\documentclass[english]{article}
\usepackage[T1]{fontenc}
\usepackage[latin9]{inputenc}
\usepackage{geometry}
\usepackage{float}
\usepackage{mathtools}
\usepackage{amsmath}
\usepackage{amsthm}
\usepackage{amssymb}
\usepackage{graphicx}

\makeatletter

\providecommand{\tabularnewline}{\\}
\floatstyle{ruled}
\newfloat{algorithm}{tbp}{loa}
\providecommand{\algorithmname}{Algorithm}
\floatname{algorithm}{\protect\algorithmname}

\theoremstyle{plain}
\newtheorem{thm}{\protect\theoremname}
\theoremstyle{plain}
\newtheorem{lem}[thm]{\protect\lemmaname}
\theoremstyle{definition}
\newtheorem{example}[thm]{\protect\examplename}
\theoremstyle{remark}
\newtheorem{rem}[thm]{\protect\remarkname}

\makeatother

\usepackage{babel}
\providecommand{\examplename}{Example}
\providecommand{\lemmaname}{Lemma}
\providecommand{\remarkname}{Remark}
\providecommand{\theoremname}{Theorem}

\begin{document}
\title{A principle feature analysis
\date{}
\author{ Tim Breitenbach{\thanks{  Biozentrum, Universit\"at W\"urzburg, Am Hubland, 97074 W\"urzburg, Germany; tim.breitenbach@mathematik.uni-wuerzburg.de}} \; Lauritz Rasbach{\thanks{ Department of Computer Science Distributed Systems Programming, Technische Universit\"at Darmstadt, Hochschulstra{\ss}e 10, 64289 Darmstadt, Germany; rasbachlauritz@googlemail.com}}\,\; Chunguang Liang{\thanks{  Biozentrum, Universit\"at W\"urzburg, Am Hubland, 97074 W\"urzburg, Germany; liang@biozentrum.uni-wuerzburg.de}} \; Patrick Jahnke{\thanks{Department of Computer Science Distributed Systems Programming, Technische Universit\"at Darmstadt, Hochschulstra{\ss}e 10, 64289 Darmstadt, Germany; jahnke@dsp.tu-darmstadt.de}}\; }}
\maketitle
\begin{abstract}
A key task of data science is to identify relevant features linked
to certain output variables that are supposed to be modeled or predicted.
To obtain a small but meaningful model, it is important to find stochastically
independent variables capturing all the information necessary to model
or predict the output variables sufficiently. Therefore, we introduce
in this work a framework to detect linear and non-linear dependencies
between different features. As we will show, features that are actually
functions of other features do not represent further information.
Consequently, a model reduction neglecting such features conserves
the relevant information, reduces noise and thus improves the quality
of the model. Furthermore, a smaller model makes it easier to adopt
a model of a given system. In addition, the approach structures dependencies
within all the considered features. This provides advantages for classical
modeling starting from regression ranging to differential equations
and for machine learning. 

To show the generality and applicability of the presented framework
2154 features of a data center are measured and a model for classification
for faulty and non-faulty states of the data center is set up. This
number of features is automatically reduced by the framework to 161
features. The prediction accuracy for the reduced model even improves
compared to the model trained on the total number of features. A second
example is the analysis of a gene expression data set where from 9513
genes 9 genes are extracted from whose expression levels two cell
clusters of macrophages can be distinguished.
\end{abstract}

\section{Introduction}

Data driven modeling and data driven decision making are rational
ways of exploiting information contained in data to turn it into knowledge.
The knowledge obtained from a set of data in turn may help to understand
processes better from which the data is measured. The better understanding
of the dynamics in a system may be used to steer or influence the
observed process such that we obtain our desired output. However,
the more accurate our demands become concerning the difference between
the predicted and the real outcome of a process or experiment, respectively,
the more variables need to be measured. Depending on this accuracy
requirement, measuring variables that describe small effects on the
outcome cannot be neglected. A growing amount of data is challenging
with respect to storage and processing capabilities. Furthermore,
the analysis of the data needs more effort to obtain the desired insights
into the system's relations due to the growing degrees of freedom.
The research field of molecular biology illustrates this development.
In the beginning, the nineteenth century, only light microscopes were
available to observe cells, e.g. via staining. Nowadays there are
technologies available like single cell RNA sequencing \cite{wilhelm2009rna,stuart2019integrative,tang2019single}
allowing to measure the gene expression levels of a bulk of cells
resolved at a level of an individual cell. This allows for modeling
the inner mechanisms of cells very detailed once we are able to see
the relations in the amount of data. Due to the amount of data it
is very purposeful to use computational methods for the analysis of
the data \cite{chen2019single}. 

The aim of this work is to provide a framework including an algorithmic
pipeline for the task of finding the relevant features for modeling
and prediction by identifying features that are functions of other
features. Furthermore, the relations in the considered data set are
systematically analyzed by structuring the dependencies within all
the data set's features and quantities. From this analysis the building
of reasonable and purposeful models of the underlying processes may
start. 

The key issue with data sets is that not all measured features are
independent of each other. Consequently, measuring more features may
not increase the amount of information contained in the data set with
respect to the modeling or prediction task while increasing only the
size of the data set. The information which features are mutually
independent is a valuable insight for several reasons. On the one
hand, we can build a mechanistic model starting with independent features
as input variables and step by step model other features as functions
of them. For example, models can be built with differential equations
or by fitting functions with regression. On the other hand, taking
only the variables with the significant information helps to reduce
the curse of dimensionality. Consequently, the degrees of freedom
are reduced and e.g. a neuronal network is trained with the compressed
information. Concentrating the information ensures that the prediction
power of each selected feature clearly sticks out through the noise
of the data set. Thus the concentration improves the signal-to-noise-ratio
and the prediction accuracy compared to case where no selection of
the features takes place. The presented framework results in smaller
models that still carry the relevant information. In particular, the
size of neuronal networks is reduced which may simplify their analysis
concerning explainable AI \cite{samek2019explainable}. Moreover,
there are many methods \cite{ribeiro2016should,vstrumbelj2014explaining,shrikumar2017learning,datta2016algorithmic,lipovetsky2001analysis,lundberg2017unified}
providing explanations of the output of black-box machine learning
models. A machine learning model with an as small number of independent
input variables as possible will enhance their performance as well
with respect to computational issues and providing an easy but meaningful
explanation.

In the following, we list the main advantages of our framework.
\begin{itemize}
\item Applicable for linear and non-linear relations between the features
\item Model reduction based on the original features without loss of information
preserving prediction quality 
\item Increasing performance due to focussing on the relevant features
\item Structures data sets and relations between features
\end{itemize}
The advantages listed above are achieved by the combination of a statistical
test of independence for two features and a minimal cut algorithm
from graph theory which clearly identifies structures being typical
for features that are functions of other features. The presented framework
consists of the combination of the following two building blocks.
A chi-square test \cite{greenwood1996guide} testing for independence
of two features and the minimal cut decomposition from graph theory
\cite{esfahanian2013connectivity}. The information about the independence
of two features is modeled into an undirected graph, called dependency
graph, where the nodes represent the features. There is an edge between
two nodes if the corresponding features are not independent of each
other. More specific, the values taken by each feature are not independent
of each other. The second building block is a minimal cut algorithm
that dissects that graph iteratively into disjunct subgraphs meaning
that the removed nodes correspond to features that are not independent
of the left mutually independent subgraphs. Moreover, given the values
of the corresponding features of such a subgraph we cannot infer the
values of the features of the other subgraphs. However, given the
values of the features corresponding to all the disjunct subgraphs,
we may infer the vales of the removed features. Iteratively we construct
a set of independent features. The presented framework does not exclusively
require the chi-square test. Any test that measures the dependency
between two random variables can be used to generate such a graph
defined above. The key point is to use a minimal cut algorithm to
dissect the corresponding graph since this procedure identifies typical
structures that features form when they are a function of other (stochastically
independent) features. In this context features are also called variables
or arguments of a function.

Once we have this set of independent features of which the other features
are not independent, we can further process this set by choosing only
variables of which output variables are not independent. These output
variables are the values that we would like to calculate given a set
of input variables. The connection between these input variables and
the output variables, i.e. the model that connects these variables,
can be generated with several methods. These can be machine learning
models like neuronal networks, support vector machines or decision
trees that learn the relations of this independent variables to the
output variables. Furthermore, we can analyze these dependencies by
fitting functions by regression models, for example to get a deep
understanding of how the variables are connected. Another kind of
mechanistic modeling is with partial or ordinary differential equations.
With these type of equations we model the variation of the output
variables depending on the input variables with respect to space or
time. 

An application of ordinary differential equations where we have many
possible input variables and need to find out the relevant input variables
to start from is the modeling of gene regulatory networks like in
\cite{di2007dynamic,karl2013jimena}. In these networks, the regulation
of genes by the expression level of other genes is modeled. In this
field, our framework can help to identify the topology of the network
that models which gene activates or inhibits the expression of other
genes. The variables in the scenario of gene regulatory networks represent
the expression levels of the corresponding genes, the transcription
levels of RNA or translation levels of proteins. We can identify the
basic genes or proteins from where we can start to generate the network.

In Section \ref{sec:The-principle-variable}, we introduce the framework
of our principle feature analysis (PFA) and discuss the corresponding
algorithms.

In Section \ref{sec:An-application-of}, we apply the framework to
filter out the relevant features, called metrics, of a data center
environment to classify if a current measurement of the metrics implicates
that the corresponding data server is in an error state or if the
data server is in a normal working status. Furthermore the presented
method is evaluated by comparing it to other related methods among
other things.

A further field of application of the presented framework is in bioinformatics.
In particular, the analysis of gene expression data to find out the
significantly expressed genes corresponding to different cell states.
For example if one labels cells with pathological and physiological
according to the cell's state, the analysis with the presented framework
may provide the genes that are responsible for making the difference
in the behavior of a cell to be physiological or pathological. By
the PFA one may identify the signal cascades causing genes or proteins
which are relevant in a tumor setting for example. We demonstrate
the extraction of relevant genes to differ two cell clusters of macrophages
in Section \ref{sec:Application-of-the}.

In Section \ref{sec:Related-methods}, we relate the PFA to existing
methods. A Discussion about the chi-square test and a Conclusion complete
this work.

In the Appendix, we give technical details about dealing with features
with a continuous co-domain and discuss requirements to apply the
chi-square test for our framework.

\section{The principle feature analysis (PFA)\label{sec:The-principle-variable}}

In this section, we describe the principle feature analysis. The description
includes all the necessary definitions, algorithms, examples to illustrate
the analysis and a theoretical result. We start with describing the
basic idea and subsequent we explain the framework in detail. 

\begin{figure}[H]
\center\includegraphics[scale=0.5]{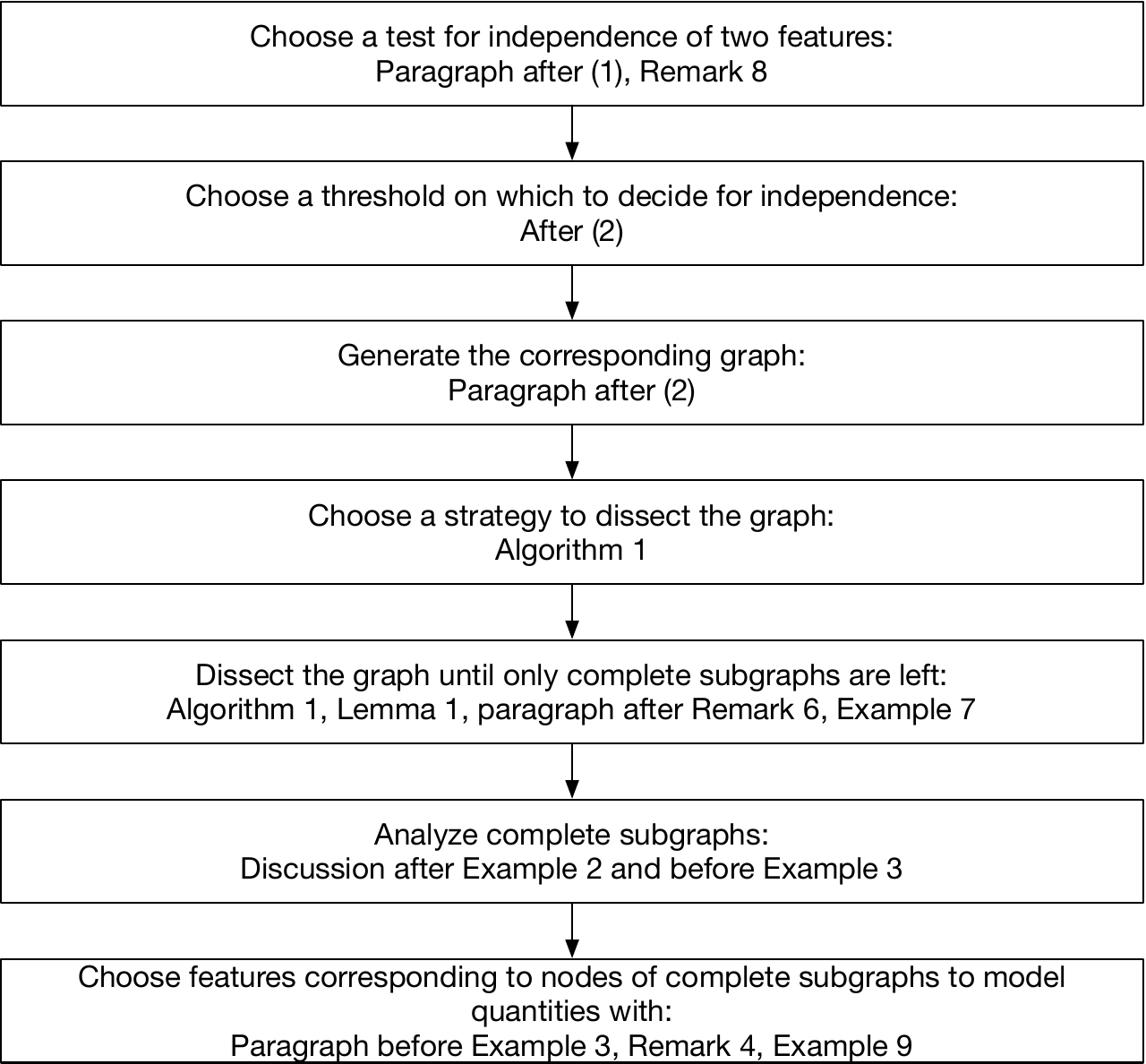}\caption{\label{fig:Workflow-of-Section}Workflow of Section \ref{sec:The-principle-variable}
with sites in the text where to find information about the topic.}

\end{figure}

The idea in a nutshell: our first step for analyzing the relation
between the features is to test if two features are stochastically
independent. If the features are stochastically independent, the value
of one feature does not influence the value of the other feature.
On the other hand, features that are a function of other features
are not independent of these input features. The key issue is to identify
such structures that functions form with their input variables. Thus
we distinguish between features that are functions and features that
are the input variables or arguments of functions. The independency
is evaluated based on the result of a suitable statistical test with
which we test the hypothesis that two features are stochastically
independent. The suitable stochastic test has to provide a probability
for the test's result of the investigated instance of values of the
corresponding features given the hypothesis is true. Since with a
measurement we can only pick a specific instance of a random experiment,
we have to define a measure on which we decide to reject the hypothesis.
For this purpose, we have to define a threshold, we call it the level
of significance. If the probability of the test is below this threshold,
we consider it too unlikely that the hypothesis holds in the considered
tested instance and we rather assume that the opposite of the hypothesis
is correct, i.e. the features are stochastically not independent.
From the information about the independency of each two features,
we generate a graph where each node represents a feature. The binary
result of the test of independency is encoded with unweighted and
undirected edges. An edge between two nodes indicates that the corresponding
features are statistically not independent based on our predefined
level of significance. If two features are stochastically independent,
then these two features do not influence each other. Consequently,
there is no functional relation between such features with which we
could calculate the value of one of these features given the value
of the other one in a measurement. In contrast, features which are
a function of other features do stochastically depend on them since
the outcome of the function is influenced by the values of the input
features. The nodes of independent features can be connected via paths
over nodes of features that are a function of these mutually independent
features. The crucial observation is now that features that are functions
of other features are linkers of disconnected nodes or subgraphs,
respectively. Removing such linker nodes from the graph may correspond
to identifying features that are functions of other features. As we
see later, any method that finds a set of nodes of minimal cardinality
such that the remaining graph consists of at least two disjunct subgraphs
is suitable for identifying the linker nodes. Moreover, the minimality
of the set will ensure that no independent feature is removed. We
repeat dissecting the graph by removing sets of minimal cardinality
until only complete subgraphs are left, i.e. subgraphs in which each
node is connected by an edge with any other node of this subgraph.
The features of these resulting subgraphs are considered as the input
features from where the modeling can start and construct the dependencies
to interesting features that correspond to nodes that have been removed.
In addition, from the input features we can identify these features
on which a considered quantity, which is supposed to be modeled, depends
with a further suitable statistical test before we start the modeling.
In Figure \ref{fig:Workflow-of-Section}, we summarize the workflow
and give sites in this section where the topic is investigated.

In the following, we describe the framework in detail. Let a set of
$n\in\mathbb{N}$ random variables be denoted with $\tilde{X}=\left\{ x_{1},...,x_{n}\right\} $
where each random variable is denoted with $x_{i}:\Omega\to\mathbb{R}$,
$\omega\mapsto x_{i}\left(\omega\right)$ for all $i\in\left\{ 1,...,n\right\} $
with $\Omega$ a measurable space. We refer to \cite[Chapter 1]{klenke2013probability}
for basic definitions of the topic of stochastics. Furthermore, let
$Y=\left\{ y_{1},...,y_{m}\right\} $, $m\in\mathbb{N}$ be a set
of output random variables where it holds $y_{j}:\Omega\to\mathbb{R}$,
$\omega\mapsto y_{j}\left(\omega\right)$ for all $j\in\left\{ 1,...,m\right\} $.
In this framework, each considered feature is modeled with a random
variable in the sense that the value that a feature takes is a measurement
of an outcome of a random experiment. In the following the features
will be denoted with random variables. 

The subset of $\tilde{X}$ of all principle variables, i.e. the variables
corresponding to the nodes of the resulting complete subgraphs, is
denoted with $X'\subseteq\tilde{X}$. The subset of $X'$ of which
variables in $Y$ are not independent is denoted with $X\subseteq X'$. 

We recall the concept of stochastically independent random variables,
see also \cite[Chapter 2]{klenke2013probability}, since this is an
essential ingredient of the presented work. We remark that the framework
is not limited to the concept of stochastic independence. Any measure
that describes the relation between two variables to define what the
values of one variable tells about the value of the other one can
be used. Let us consider two random variables $A:\Omega\to Z_{A}$,
$\omega\mapsto A\left(\omega\right)$ and $B:\Omega\to Z_{B}$, $\omega\mapsto B\left(\omega\right)$
where $Z_{A}$ is a set of $k\in\mathbb{N}$ elements that $A$ can
take and $Z_{B}$ is a set of $l\in\mathbb{N}$ elements that $B$
can take. The elements of $Z_{A}$ and $Z_{B}$ are called events.
We say that two variables are stochastically independent if the equation
\begin{equation}
P\left(A=z_{A}^{i}\mathrm{\ and\ }B=z_{B}^{j}\right)=P\left(A=z_{A}^{i}\right)P\left(B=z_{B}^{j}\right)\label{eq: independent random varaiables}
\end{equation}
is fulfilled with $z_{A}^{i}\in Z_{A}$, $z_{B}^{j}\in Z_{B}$ for
all $i\in\left\{ 1,...,k\right\} $ and $j\in\left\{ 1,...,l\right\} $
where $P$ is the function that gives the probability for the random
variables taking the corresponding event as denoted with the argument
of $P$ in the brackets. The function $P$ is called the probability
function. We remark that if the set of variables $\tilde{X}$ contains
variables with a continuous co-domain, i.e. the range of values a
variable can take, the co-domain of such a variable needs to be discretized
for applying (\ref{eq: independent random varaiables}). A possible
implementation is shown in the appendix, see Algorithm \ref{alg:Discretize-the-co-domain}.

In the following, we discuss the test of independence of two random
variables from $\tilde{X}$. For this purpose, it is tested if these
random variables fulfill (\ref{eq: independent random varaiables}).
Due to the finiteness of the number of measurements and the possibly
non-deterministic dynamics of a system, we cannot expect that our
measured distribution of our random variables exactly matches its
real distribution according to which each random variable distributes
its outcomes. Consequently, both sides of (\ref{eq: independent random varaiables})
will not exactly be equal even in the case where the real distribution
of the random variables, which we actually mostly do not know, fulfill
(\ref{eq: independent random varaiables}) perfectly. However, we
can test if the equality holds on a level of significance which means
that it can be tested how likely it is to obtain such different values
of the left and right hand-side of (\ref{eq: independent random varaiables})
given that both random variables are independent. In this work, a
chi-square test is applied as follows. Both sides of (\ref{eq: independent random varaiables})
can be each considered as a discrete probability distribution function
with regard to the domain $i\in\left\{ 1,...,k\right\} $ and $j\in\left\{ 1,...,l\right\} $.
Before we go ahead, we write (\ref{eq: independent random varaiables})
into an equivalent form where the frequencies of the events are considered.
We multiply (\ref{eq: independent random varaiables}) by the total
number of measurements $N$ for all $i\in\left\{ 1,...,k\right\} $
and $j\in\left\{ 1,...,l\right\} $. Then the observed frequency for
the event $A=z_{A}^{i}\mathrm{\ and\ }B=z_{B}^{j}$ is given by 
\[
f_{O}:\left\{ 1,...,k\right\} \times\left\{ 1,...,l\right\} \to\mathbb{R},\ \left(i,j\right)\mapsto f_{O}\left(i,j\right)=P\left(A=z_{A}^{i}\mathrm{\ and\ }B=z_{B}^{j}\right)\cdot N
\]
and the expected frequency of this event is given by
\[
f_{E}:\left\{ 1,...,k\right\} \times\left\{ 1,...,l\right\} \to\mathbb{R},\ \left(i,j\right)\mapsto f_{E}\left(i,j\right)=P\left(A=z_{A}^{i}\right)P\left(B=z_{B}^{j}\right)\cdot N
\]
assuming that the $A$ and $B$ are independent, meaning the outcome
of one random variable does not influence the outcome of the other
one. Since (\ref{eq: independent random varaiables}) has to hold
for all $i\in\left\{ 1,...,k\right\} $ and $j\in\left\{ 1,...,l\right\} $,
we equivalently test if both distributions with the corresponding
frequencies of the events are the same. For this purpose, the measure
\begin{equation}
\chi^{2}\coloneqq\sum_{i=1}^{k}\sum_{j=1}^{l}\frac{\left(f_{O}\left(i,j\right)-f_{E}\left(i,j\right)\right)^{2}}{f_{E}\left(i,j\right)}\label{eq:chi-square def}
\end{equation}
is used to test, if the observed distribution of the frequency $f_{O}$
equals the expected distribution of the frequency $f_{E}$ assuming
the independence of the two random variables $A$ and $B$. If both
variables are stochastically independent and thus both sides of (\ref{eq: independent random varaiables})
are supposed to be equal, the measure $\chi^{2}$ is supposed to be
small. Due to the finite number of measurements and the non-deterministic
dynamic of the random variables, it may probably happen, that the
outcome of our measurement process is such that, although the random
variables are independent and thus (\ref{eq: independent random varaiables})
holds for the real distributions, the measure $\chi^{2}$ is greater
than zero. In this case, we need to decide on a predefined level of
significance $\alpha>0$ if we consider $A$ and $B$ as independent
random variables though. In the case of independent random variables,
$\chi^{2}>0$ is considered as a result of random fluctuations. If
we define 
\begin{equation}
\chi_{ij}\coloneqq\frac{f_{O}\left(i,j\right)-f_{E}\left(i,j\right)}{\sqrt{f_{E}\left(i,j\right)}},\label{eq: def chi_ij}
\end{equation}
for any $i\in\left\{ 1,...,k\right\} $ and $j\in\left\{ 1,...,l\right\} $,
then $\chi^{2}$ is chi-square distributed if each $\chi_{ij}$ is
normally distributed with the expectation zero and the variance one
and mutually independent for all $i\in\left\{ 1,...,k\right\} $ and
$j\in\left\{ 1,...,l\right\} $, cf. \cite[Chapter 18]{cramer1999mathematical}.
Based on the level of significance $\alpha$, we can decide if our
hypothesis that $A$ and $B$ are independent has to be rejected by
considering how likely it is to obtain the calculated $\chi^{2}$
values assuming $A$ and $B$ are independent. In the appendix, we
discuss conditions such that each $\chi_{ij}$ can (approximately)
be considered a normally distributed random variable with the expectation
zero and the variance one for any $i\in\left\{ 1,...,k\right\} $
and $j\in\left\{ 1,...,l\right\} $. The mutual independence of $\chi_{ij}$
for all $i\in\left\{ 1,...,k\right\} $ and $j\in\left\{ 1,...,l\right\} $
is discussed as well.

Next, we define a graph from the information if two random variables
are independent based on the considered data. We call such a graph
a dependency graph. The information of independency of two random
variables can be translated into an adjacency matrix $M$ where an
entry of value one means that the corresponding random variables are
mutually not independent and the value zero means that they are mutually
independent. Formally for the case where we analyze the set $\tilde{X}$,
we define 
\begin{equation}
M_{ij}\coloneqq\begin{cases}
1 & \mathrm{\ if\ }x_{i}\mathrm{\ is\ not\ independent\ of\ }x_{j}\\
0 & \mathrm{\ if\ }x_{i}\mathrm{\ is\ independent\ of\ }x_{j}
\end{cases}\label{eq:adja ma}
\end{equation}
for all $i,j\in\left\{ 1,...,n\right\} $. Analogously, we can define
an adjacency matrix for the case where we analyze the independency
of random variables to output variables in the set $Y$ where one
index of the adjacency matrix represents one input variable and the
other index represents an output variable. The adjacency matrix $M$
is symmetric since the roles of the variables in (\ref{eq: independent random varaiables})
can be interchanged. 

Next, we discuss some special cases of stochastic independence. If
a random variable is constant, we choose $A$, then (\ref{eq: independent random varaiables})
is always fulfilled since $P\left(A=z_{A}^{i}\right)=1$ for $i\in\left\{ 1\right\} $
and $P\left(A=z_{A}^{i}\mathrm{\ and\ }B=z_{B}^{j}\right)=P\left(B=z_{B}^{j}\right)$
for all $i\in\left\{ 1\right\} $ and $j\in\left\{ 1,...,l\right\} $
because all the events where $B=z_{B}^{j}$ do not branch out due
to $A$ being constant. Consequently, $A$ and $B$ are independent
in the case where one variable is constant. As a special case any
constant random variable is independent to itself according to the
definition given by (\ref{eq: independent random varaiables}). In
contrast is a non-constant random variable to itself. If two random
variables are each non-constant and identical, we denote both with
$A$, then (\ref{eq: independent random varaiables}) can never be
fulfilled since $P\left(A=z_{A}^{i}\right)<1$ for one $i\in\left\{ 1,...,n\right\} $.
Consequently, considering (\ref{eq: independent random varaiables})
for this $i$ where $P\left(A=z_{A}^{i}\right)<1$, we have that 
\[
P\left(A=z_{A}^{i}\right)=P\left(A=z_{A}^{i}\mathrm{\ and\ }A=z_{A}^{i}\right)\neq P\left(A=z_{A}^{i}\right)P\left(A=z_{A}^{i}\right)=P\left(A=z_{A}^{i}\right)^{2}
\]
since $P\left(A=z_{A}^{i}\right)>0$ for all $i\in\left\{ 1,...,k\right\} $
and the counts where $A=z_{A}^{i}$ equals $A=z_{A}^{i}\mathrm{\ and\ }A=z_{A}^{i}$
because we have no sub cases for all the events where $A=z_{A}^{i}$.
Intuitively, this makes sense, since once we know the value of one
random variable we can predict the values of all identical random
variables by mapping the  known value by the identity function. We
remark that for the intention of the work, we are interested in the
links of one variable to other ones. For convenience, we consequently
define $M_{ii}\coloneqq0$ for all $i\in\left\{ 1,...,n\right\} $.
For our consideration, as we discussed so far, the relevant information
is in the upper triangular matrix of $M$ without the diagonal. 

The mutual stochastic independency of the elements of $\tilde{X}$
is modeled by an adjacency matrix which can be visualized by a corresponding
undirected graph $G$ due to the symmetry of $M$. In this graph $G$,
a node represents a random variable of $\tilde{X}$ and an edge between
two nodes represents that the two corresponding random variables are
mutually not independent of each other. Now, we explain how this graph
can be further analyzed in order to obtain a set of independent variables
$X'$ of which all the random variables $\tilde{X}$ are not independent.
The elements of $X'$ can then be taken as the starting point or input
variables, respectively, for any kind of modeling like (non-)linear
regressions, differential equations or neuronal networks in order
to quantitatively describe the dependencies existing for the random
variables of $\tilde{X}$.

In the following, we give and discuss the algorithm to extract $X'$
from $\tilde{X}$. The graph is iteratively dissected into disconnected
subgraphs by removing a set of nodes of minimal cardinality until
only complete graphs are left. The removed nodes are not independent
of the subgraphs being independent to the other subgraphs, i.e. each
variable of a corresponding node of a subgraph is independent to all
the other variables represented by nodes of the other subgraphs. The
algorithm is given in Algorithm \ref{alg:Graph-dissection-algorithm}.

\begin{algorithm}[H]
\begin{enumerate}
\item Sort $G$ into subgraphs: $G_{c}$ are the complete subgraphs and
$G'=G\backslash G_{c}$ are the subgraphs that are not complete.
\item For $g\in G'$
\begin{enumerate}
\item Select a set of nodes $S$ of $g$ of minimal cardinality such that
$g$ is decomposed into disjunct subgraphs when removing $S$.
\item Remove $g$ from $G'$.
\item Remove $S$ from $g$. Name the resulting graph $g$.
\item For any element in $g$: 
\begin{enumerate}
\item If $\tilde{g}\in g$ is complete: Add $\tilde{g}$ to $G_{c}$
\item If $\tilde{g}\in g$ is not complete: Add $\tilde{g}$ to $G'$
\end{enumerate}
\item Stop if $G'$ is empty.
\end{enumerate}
\end{enumerate}
\caption{\label{alg:Graph-dissection-algorithm}Graph dissection algorithm}
\end{algorithm}

Next, we state and prove a lemma that says that Algorithm \ref{alg:Graph-dissection-algorithm}
is well-defined.
\begin{lem}
\label{lem:Algorithm--is}Algorithm \ref{alg:Graph-dissection-algorithm}
is well-defined. That means the for-loop in Step 2 stops after finitely
many iterations. The subgraphs of $G_{c}$ are mutually disconnected
and there is a path in $G$ to any node of $G\backslash G_{c}$ starting
in a subgraph of $G_{c}$. Furthermore, there is no random variable
corresponding to a node of $G$ that is stochastically independent
of all the random variables represented by the nodes of $G_{c}$.
\end{lem}

\begin{proof}
If $G$ is a complete subgraph or consists of only disconnected complete
subgraphs, i.e. $G_{c}=G$, then the statement is true. If a subgraph
of $G$ is not complete, then there are at least two nodes that are
not connected, i.e. there are two independent random variables. Let
$n'$ denote the number of nodes of this incomplete subgraph. Then
there exists a set of $n'-2$ nodes that can be removed such that
this incomplete subgraph decomposes in two complete subgraphs. Since
there is a set of finitely many elements, there also exists a set
of minimal cardinality that decomposes the graph upon removing it.
The cardinality of this minimal set is at least one since we have
an incomplete graph with two disconnected nodes and consequently we
need at least a third node connected to these two nodes. Otherwise
it would be a graph consisting of two disconnected complete subgraphs.
Thus, we have proved that we stop after finitely many steps, since
we at least remove one node from a set with finitely many nodes as
long as our initial graph is not decomposed into disconnected complete
subgraphs.

Since the subgraphs of $G_{c}$ are disjoint per construction, we
next prove that there is a path to any node of $G\backslash G_{c}$
in $G$ starting form a subgraph of $G_{c}$. For this purpose, we
consider two non-connected (non-neighbored) nodes $s\in G$ and $t\in G$
which always exists if the graph is not complete according to the
discussion in the previous paragraph. We remove a set of nodes of
minimal cardinality $V$ from the graph such that these two nodes
are each in a subgraph disjoint from each other. Then all the removed
nodes were connected to both remaining graphs, which are graph $S$
which contains $s$ and the graph $T$ that contains $t$. That means
there exists a path to each ever removed nodes from a remaining node.
Iteratively we can consequently find a path from a node of $G_{c}$
to a node of $G$.

Next, we prove that in each iteration, we do not remove a node corresponding
to a random variable that is stochastically independent of the random
variables corresponding to the remaining nodes. Then, we have that
there is no random variable corresponding to a node of $G$ that is
independent of all the random variables represented by the nodes of
$G_{c}$. Let $V$ be a set of minimal cardinality that dissects a
graph into a subgraph $S$ and a subgraph $T$ when being removed.
If there was a node $e\in V$ which was not (directly) connected to
a node of $S$ and $T$ or in other words neighboring a node of $S$
and $T$, i.e. the corresponding random variable is independent of
all the random variables represented by the nodes of $S$ and $T$,
then this would be a contradiction to the requirement that the set
of nodes $V$ was of minimal cardinality to separate $S$ and $T$
since we could remove only $V\backslash\left\{ e\right\} $ which
would still separate $S$ and $T$.
\end{proof}
With an illustrating example, we introduce how Algorithm \ref{alg:Graph-dissection-algorithm}
works and motivate its usefulness. In particular, the example shows
how Algorithm \ref{alg:Graph-dissection-algorithm} identifies structures
that variables generate that are a function of other variables. Furthermore,
we compare our framework to a naive approach for identifying principle
variables.
\begin{example}
\label{exa:We-choose-....}We choose three pairwise stochastically
independent random variables $x_{1}$, $x_{2}$, $x_{3}$ and set
$x_{4}=2v_{1}v_{2}v_{3}$, $x_{5}=v_{1}v_{2}$. The graph's adjacency
matrix is generated as defined in (\ref{eq:adja ma}) and the graph
is visualized in Figure \ref{fig:A-dependency-graph}. Now, the task
is to select a set of variables from which we can reconstruct the
dependencies from all the variables $x_{1}$, ...., $x_{5}$. 

Before the procedure of Algorithm \ref{alg:Graph-dissection-algorithm}
is exemplified, a naive strategy is illustrated that might be considered
for the task. The naive approach for identifying principle variables
starts with choosing a node randomly. Then, we iterate over all the
other variables and include any variable into our set of input variables
that is stochastically independent of all variables that are contained
in our current set of input variables. By going through the possibilities
to choose an initial node, we see that the result in this case depends
on the initially chosen variable. If we take $x_{4}$ for example,
no other variable will be chosen and we are not able to construct
the values of the other random variables. Next, we see that the result
of Algorithm \ref{alg:Graph-dissection-algorithm} is unique in this
case.

By Algorithm \ref{alg:Graph-dissection-algorithm}, we uniquely choose
the set $\left\{ x_{1},x_{2},x_{3}\right\} $ as input variables which
allows us to construct all the values of the remaining random variables.
To illustrate the Algorithm \ref{alg:Graph-dissection-algorithm},
we describe how it proceeds in this example. The first set of nodes
that is removed is $\left\{ x_{4}\right\} $, since there is no other
one element set that dissects the graph upon removing it. Then only
the graph consisting of $x_{1}$, $x_{2}$ and $x_{5}$ is further
processed, since the graph consisting of the single node $x_{3}$
is complete. In the first graph, since it is not complete, the only
node to remove to obtain two dissected complete subgraphs is node
$x_{5}$. From the result $\left\{ x_{1},x_{2},x_{3}\right\} $, we
can exactly construct all the other two nodes. For example, we can
find the functional dependencies by regression methods staring from
the variables $x_{1}$, $x_{2}$ and $x_{3}$ as the domain variables.
The described procedure is also shown in Figure \ref{fig:Demonstration-of-Algorithm}

This example demonstrates that Algorithm \ref{alg:Graph-dissection-algorithm}
might be a better choice for identifying variables by which we can
model all the remaining variables or predict the output of a system,
respectively, than just using the naive approach introduced two paragraphs
before. The reason is that stochastically independent variables are
not (directly) connected to each other. The variables or functions,
respectively, that are not independent of several of these independent
variables link them together. Consequently these variables representing
functions are likely to be removed by Algorithm \ref{alg:Graph-dissection-algorithm}
in order to obtain disconnected subgraphs. This leaves the real input
variables left as disconnected subgraphs.

\begin{figure}[H]
\center\includegraphics[scale=0.5]{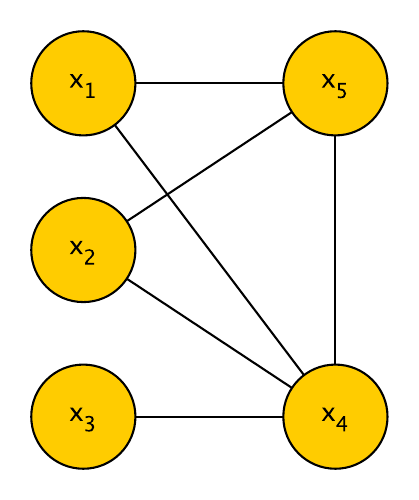}

\caption{\label{fig:A-dependency-graph}A dependency graph for Example \ref{exa:We-choose-....}.}

\end{figure}

\begin{figure}[H]
\center\includegraphics[scale=0.4]{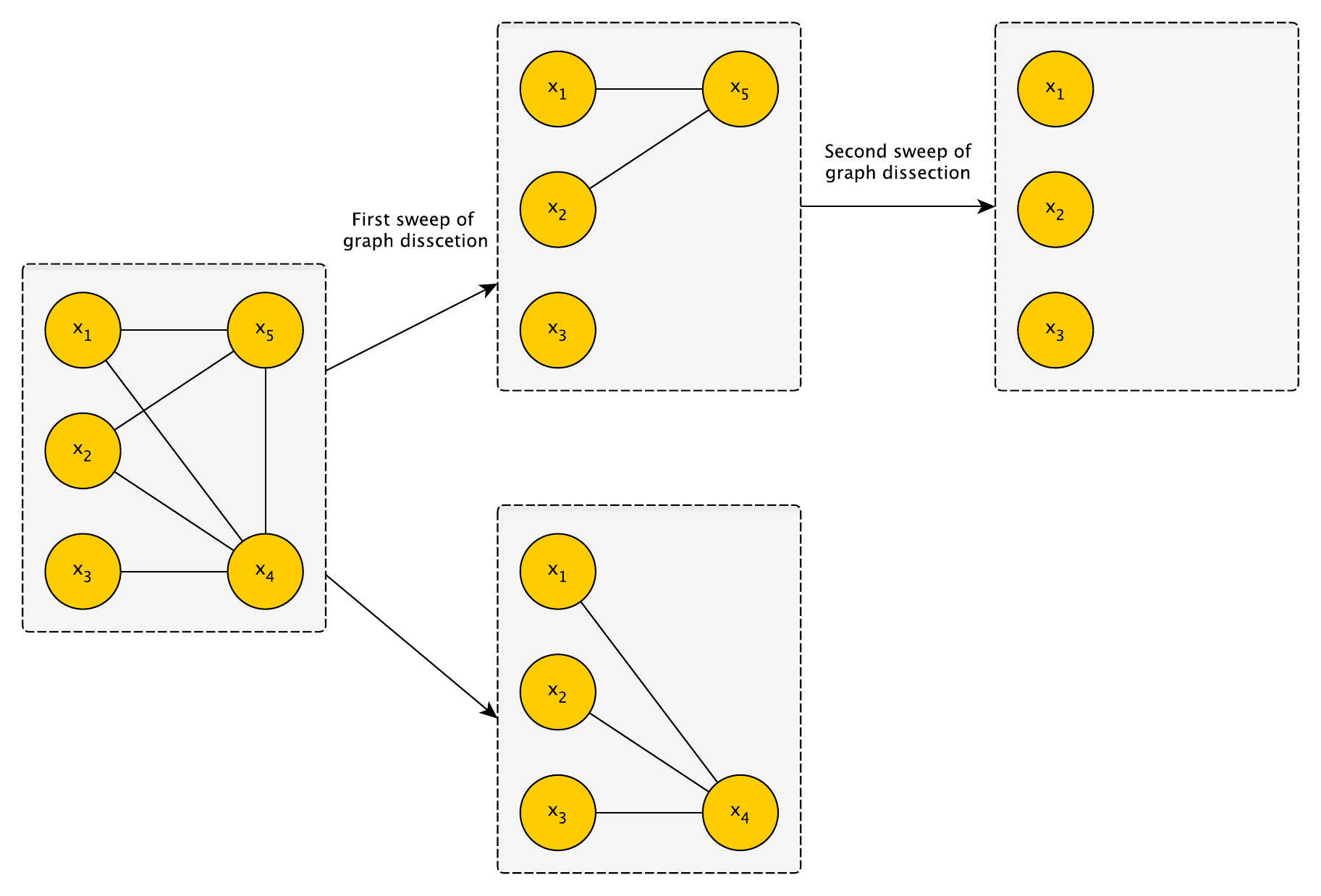}

\caption{\label{fig:Demonstration-of-Algorithm}Demonstration of Algorithm
\ref{alg:Graph-dissection-algorithm}. The branch on top shows how
the dissection works while in the branch below we see that removing
$x_{5}$ in the first step does not dissect the graph into disjunct
subgraphs.}
\end{figure}
\end{example}

\label{exa:We-see-that}In the next step, the meaning of the complete
subgraphs that Algorithm \ref{alg:Graph-dissection-algorithm} returns
is discussed. The complete subgraphs are the input interfaces of the
dependency graph of our considered system of which quantities are
measured. These complete subgraphs may be interpreted as the influx
of information that propagates through the system and the values of
the corresponding variables determine the values of the other random
variables corresponding to the removed nodes. In the case where a
complete subgraph returned by Algorithm \ref{alg:Graph-dissection-algorithm}
consists of only one node, we can use the corresponding random variable
as an input variable for our model of the considered system. In the
case where a complete subgraph consists of at least two nodes, we
discuss some cases in the following on how to proceed.

Nodes of a complete subgraph correspond to random variables that are
pairwise not independent. There are some cases that can result in
complete subgraphs consisting of more than just one node. We discuss
them in the following.

In the first case, each variable of a complete subgraph can be described
by a single variable of this subgraph with a function. More specific,
there exists a function to describe each random variable represented
by a node of the subgraph with another random variable whose node
is an element of the considered subgraph. As an example choose the
dependency $x_{1}=x_{2}^{2}$ as a functional connection between two
random variables. Both variables are not independent of each other
and this functional dependency can be described by either $x_{1}=x_{2}^{2}$
or $x_{2}=\sqrt{|x_{1}|}$. However, the square root function is not
smooth considered as a function, where in contrast the square function
is smooth, i.e. it can be differentiated. This can be important depending
on the application of the model. For example if derivatives of the
model are needed since it is used in an optimal control problem or
optimization framework. Consequently, in the case where one node of
the complete subgraph is sufficient to represent all the values of
the other nodes by corresponding functions, the nodes of the complete
subgraph can be considered as algebraically equivalent but not analytically.

\label{The-second-case}The second case where subgraphs can be obtained
is that relevant variables are not measured. Not measuring relevant
variables can have several reasons like that one has to decide what
variables have to be recorded or measured due to limitation of storage
or other lack of performance or processing power, respectively. Another
reason is, that one was not aware that a variable could have impact
on the description of a system or an output variables and for this
reason an important variable is not recorded or measured. We illustrate
the scenario with the following example. Consider $x_{3}=x_{1}-x_{2}$
as a function of the mutually stochastically independent random variables
$x_{1}$ and $x_{2}$. Only the variable $x_{3}$ and $x_{2}$ are
measured. Since $x_{3}$ and $x_{2}$ are not independent of each
other, the nodes form a complete subgraph and thus this graph is unchanged
returned by Algorithm \ref{alg:Graph-dissection-algorithm}. In the
case that we would like to predict an output variable that only depends
on $x_{2}$, the inclusion of $x_{3}$ to the set $X$ besides $x_{2}$
would not provide a better prediction accuracy since all the necessary
and sufficient information is already contained in the variable $x_{2}$.
In this case, the node of $x_{3}$ of the complete subgraph could
be neglected. In the case that a desired output variable is not independent
of $x_{1}$ (but $x_{1}$ is not measured), we could calculate the
value of $x_{1}$ from the values of $x_{2}$ and $x_{3}$ by adding
$x_{3}$ and $x_{2}$. Alternatively, a machine learning model could
be trained and during this process the functional dependency between
$x_{2},x_{3}$ and $x_{1}$ would be implicitly learned to construct
the variable $x_{1}$ given the values of $x_{2}$ and $x_{3}$ and
then perform the prediction of the output variable. In this case the
prediction is likely to perform better, if we include $x_{2}$ and
$x_{3}$ into $X$. The benefit of our principle feature analysis
is that the analysis for modeling can be broken down to only those
subgraphs with more than one node. Only the complete subgraphs with
more than one node have to be taken care of more closely depending
on the purpose and required accuracy of the prediction or modeling.

However, although considering all the nodes of a subgraph, the variables
corresponding to that subgraph might not contain sufficient information
to make a satisfactory prediction since too many variables are not
measured that are important for a sufficient prediction or sufficiently
accurate model. An example can be where $x_{4}=x_{1}+x_{2}-x_{3}$.
If only $x_{4}$ and $x_{1}$ are measured, the nodes of $x_{4}$
and $x_{1}$ form a complete subgraph and we could for example not
make a good prediction if an output variable explicitly needs the
value of $x_{2}$, like $y=x_{2}+x_{3}$. However, if the output variable
would be like $y=x_{2}-x_{3}$ we could make a good prediction since
by subtracting $x_{1}$ from $x_{4}$ would perfectly result in the
difference of $x_{2}-x_{3}$. Consequently, if the prediction is not
sufficient it may be a hint to measure more different variables, i.e.
that the relevant information for the prediction or modeling is not
included in the current measurements.

Summarizing, our framework can accelerate modeling since from all
the variables of the total data set, which are initially all possible
input variables, we extract a subset of variables. From this subset
only the sets of variables corresponding to complete subgraphs with
more than one node need to be further investigated to make a set of
input variables from where a detailed analysis of the dynamics of
the considered system can start. This analysis can be used to find
the main causes or rules to predict an output variable of a system
that we are interested to forecast.

In the third case, a complete subgraph can consist of many nodes.
Having many nodes in a complete subgraph can be a hint that more variables
should be measured. The example where $x_{3}=x_{1}\cdot x_{2}$ with
the mutually stochastically independent random variables $x_{1}$
and $x_{2}$ shows that if we do not measure $x_{1}$ for example,
the corresponding complete subgraph that Algorithm \ref{alg:Graph-dissection-algorithm}
returns consists of the nodes for $x_{2}$ and $x_{3}$. If in addition
$x_{1}$ is measured, Algorithm \ref{alg:Graph-dissection-algorithm}
returns the two complete subgraphs $x_{1}$ and $x_{2}$ each consisting
of only one node. Consequently, measuring more variables can result
in complete subgraphs with less nodes each since more variables can
resolve the information of the considered system more accurately resulting
in more branches of the dependency graph that can be dissected.

We conclude our discussion about the complete subgraphs by showing
how to approximate exact relations of variables represented by a complete
subgraph. In other words, we analyze how much relevant information
a variable carries with regard to calculating an (output) variable
that is supposed to be modeled. The number of nodes of a complete
subgraph returned by Algorithm \ref{alg:Graph-dissection-algorithm}
can be further reduced depending on our requirement for accuracy that
we have for a model or a prediction. For example let's consider $x_{3}=x_{1}+10^{-3}\cdot x_{2}$
for the mutually independent random variables $x_{1}$ and $x_{2}$
with the same order of magnitude of values that $x_{1}$ and $x_{2}$
can take. Then formally Algorithm \ref{alg:Graph-dissection-algorithm}
returns $x_{1}$ and $x_{2}$ each as complete subgraphs. However,
depending on the level of accuracy, we can find out by analytical
investigations that already $x_{1}$ is sufficient as a model to predict
or calculate the values for $x_{3}$. The benefit of our method is,
that it reduces the variables that need to be considered for modeling
or for machine learning in a preprocessing step. The preprocessing
step with our principle variable analysis extracts only these variables
that carry the information. 

\label{In-order-to}In order to find a model to describe the output
variables in the set $Y$, it is not necessary to build a model and
describe the variables of $\tilde{X}$ by the ones in $X'$ and thus
reconstruct all the relations in $\tilde{X}$. The relevant variables
of $X'$ of which the variables of $Y$ are not independent can be
analogously identified with a chi-square test as described above in
the paragraphs about the chi-square test. In detail, for any $i\in D\subseteq\left\{ 1,...,n\right\} $
with $x_{i}\in X'$, we perform a chi-square test for any $j\in\left\{ 1,...,m\right\} $
with $y_{j}\in Y$. If for a chosen $i\in D$ for one $j\in\left\{ 1,...,m\right\} $
the result of the corresponding chi-square test is that the considered
variables are not independent, then we add $x_{i}$ to $X$ as well
as all the other random variables whose representing nodes are contained
in the complete subgraph where the node of $x_{i}$ is in due to the
reasoning above in the paragraphs about the complete subgraphs starting
on page \pageref{exa:We-see-that}. To investigate the connection
between $X'$ and $Y$ to generate $X$, we are not limited to the
chi-square test but can take any test that seems suitable to relate
the variables of $X'$ to the ones of $Y$.

The next example demonstrates that first generating $X'$ from $\tilde{X}$
with Algorithm \ref{alg:Graph-dissection-algorithm} and then make
a chi-square test to obtain $X$ does not commute with first making
a chi-square test to find all the random variables of $\tilde{X}$
not being independent of the variables of $Y$ and then apply Algorithm
\ref{alg:Graph-dissection-algorithm}.
\begin{example}
\label{exa:We-choose-the}We choose the following two mutually stochastically
independent variables $x_{1}$ and $x_{2}$. Furthermore, we choose
$x_{3}=x_{1}\cdot x_{2}$. The output variable $y$ is supposed to
be only a function of $x_{1}$, for example $y=1$ if $x_{1}$ is
equal or greater a certain threshold and $y=0$ else.

According to the presented framework first the adjacency matrix is
generated for all the random variables of the set $\tilde{X}=\left\{ x_{1},x_{2},x_{3}\right\} $
by the chi-square test. The graph looks like that the node representing
$x_{1}$ and $x_{2}$ are connected with the one of $x_{3}$ and the
nodes of $x_{1}$ and $x_{2}$ are disconnected. After applying Algorithm
\ref{alg:Graph-dissection-algorithm} to this graph, we have the set
$X'=\left\{ x_{1},x_{2}\right\} $. Applying the chi-square test to
identify the variables from $X'$ that are not independent of $y$,
we obtain $X=\left\{ x_{1}\right\} $ which is exactly the variable
that we need as an input variable to describe $y$. 

Now, we try the second procedure. If we first start with a chi-square
test to identify variables from the set $\left\{ x_{1},x_{2},x_{3}\right\} $
that are not independent of $y$, we obtain $\left\{ x_{1},x_{3}\right\} $.
The corresponding graph is complete where the node representing $x_{1}$
is connected with the node representing $x_{3}$. Applying Algorithm
\ref{alg:Graph-dissection-algorithm} returns this complete graph. 

Consequently, the result from both procedures are different and the
procedures do not commute in general.
\end{example}

Based on Example \ref{exa:We-choose-the}, the following remark can
be stated.
\begin{rem}
The process of generating $X'$ from $\tilde{X}$ and then generating
$X$ from $X'$ does not commute with applying Algorithm \ref{alg:Graph-dissection-algorithm}
to all the variables of $\tilde{X}$ that are not independent of the
output variables $Y$.
\end{rem}

In the next remark, we describe how to find the input variables to
model an arbitrary variable of $\tilde{X}$ and how to reduce an existing
model.
\begin{rem}
\label{rem:We-can-express}We can express an arbitrary random variable
$x\in\tilde{X}$ by a set of other variables based on the adjacency
matrix that is constructed during the principle feature analysis as
follows. From the row corresponding to $x$ of adjacency matrix that
is calculated for the elements of $\tilde{X}$ we can take all the
other random variables in $X'$ which are connected to $x$. As a
result there is a set of independent variables that may contain all
the information of the data set to model $x$. In addition, if desired,
variables that are connected to $x$ (see the row for $x$ in the
adjacency matrix) but are not a variable of the set $X'$ can be taken
and build a model for $x$. Using the information of the adjacency
matrix is a systematic way to identify a set of independent random
variables to model a certain random variable representing a quantity
of a system. 

Furthermore, by applying Algorithm \ref{alg:Graph-dissection-algorithm}
to a set of variables that are already used in a model, the necessary
variables that capture all the information in the data set can be
extracted and thus redundancies can be removed. However, in some cases
there are pitfalls that are discussed in the following.

For example if $x_{4}=x_{1}+x_{2}+x_{3}$ where $x_{1}$, $x_{2}$
and $x_{3}$ are mutually independent. If $x_{3}$ is not measured,
the information contained in $x_{3}$ can be recovered as long as
$x_{4}$, $x_{1}$ and $x_{3}$ are available. Applying Algorithm
\ref{alg:Graph-dissection-algorithm} to the set $\left\{ x_{1},x_{2},x_{4}\right\} $
will remove node $x_{4}$ since the corresponding variable is not
independent form the mutually independent variables $x_{1}$ and $x_{2}$.
Consequently removing a node may cause a loss of relevant information
and may result in the following. A consequence may be that modeling/prediction
with the variables identified by Algorithm \ref{alg:Graph-dissection-algorithm}
is not sufficient, i.e. the errors between the prediction of the model
and the measured data are too big for the given tolerances. This consequence
can be easily checked when reducing input variables of an existing
model by comparing the accuracy of the previous model with the accuracy
of the model with the reduced number of input variables. 

If the prediction accuracy of the model based on the new set of input
variables is still sufficiently good, the previous set of input variables
and equivalently the information can be recovered by the new set of
input variables. The link between the old and new set of input variables
consists of functional relations that compress the information necessary
for the prediction. The advantage is that already existing well working
models can be reduced resulting in a faster calculation of the output
or a better interpretability of the total model, e.g. in terms of
machine learning models.
\end{rem}

Next, we discuss how to improve the accuracy of an existing model
by adding further variables. At the same time the framework is used
to reduce redundancy from the set of input variables of the considered
model.
\begin{rem}
\label{rem:If-we-choose} Any existing model whose predictions or
calculations of output variables are considered to be not sufficient
(any more), i.e. the predicted output does not fit well to the corresponding
measured data, may be improved by adding further variables to the
set of input variables since not sufficient information is contained
within the current input variables. Consequently, supplementing the
set of the current input variables with more variables that describe
further quantities of our considered system may help to improve the
accuracy. However, using sufficiently much of the information of a
data set is just a necessary criterion for a sufficiently accurate
model. The reason is that maybe not all the information to make a
sufficient modeling is contained in our data set. Once the prediction
based on the model with our chosen input variables is sufficient,
the presented framework can be used to remove potentially existing
redundancies from the set of input variables as shown in Remark \ref{rem:We-can-express}.
\end{rem}

Remark \ref{rem:We-can-express} creates awareness of checking a model's
prediction accuracy after removing each node to possibly compress
an arbitrary set of input nodes in order not to lose information.
The same procedure holds in principle for the case where we apply
the framework to the total data set to generate a model from the scratch
where there is mostly no evidence to have all the necessary variables
in the data set for a sufficient prediction. The advantage of the
presented framework is that it contains a systematical way to analyze
when information is lost by removing a variable from the data set
to obtain a purposeful set of input variables. For example each time
a node is removed, it can be checked with a model if the prediction
of the values corresponding to the removed nodes is sufficiently good
based on the values of the variables corresponding to the remaining
nodes in the considered subgraph. A model can be generated by e.g.
a machine learning model which can take the action of an oracle answering
the question if the prediction of the removed variables based on the
remaining variables is still sufficient. If the prediction is not
sufficient, the framework thus provides hints by this step by step
analysis where to improve the data acquisition. For example, if the
prediction of the values of a variable corresponding to a removed
node is not sufficient, then removing this variable means losing information.
In order to prevent this loss, it can be thought about what further
variables of the considered system might be measured to improve the
prediction of the variable since maybe not all effects that influence
the removed variables are captured so far. However, we are aware of
the fact that if a prediction is not sufficient, it does not necessarily
have to mean that information is lost by removing a node. It can mean
that the considered model is not suited to describe the relation between
the removed and the remaining variables. For generating more confidence
on this test if information is lost when removing nodes, e.g. several
different machine learning models can be utilized. If at least one
model has a sufficient prediction accuracy, the information in the
reduced data set is adequate.

If the prediction of the output variable from the set of input variables
$X$ is not sufficient, those variables could additionally be included
into $X$ that were removed and could not have been predicted well
by the remaining variables. An improvement of the prediction can come
from the fact that not all relevant factors that influence a removed
variable are captured. By including such a removed variable whose
prediction was not good, we can include relevant information as shown
in e.g. the paragraph starting on page \pageref{The-second-case}
about complete subgraphs. However, if we have thousands of variables
this procedure may not be practicable, for example since a neuronal
network might not be practically trained on the many variables that
are given in the beginning of the procedure. In this case, it may
be easier to just measure more variables of the considered system
that are not measured so far to improve the prediction. By measuring
more and more variables sufficient information may be collected for
an accurate prediction.

In the next example, we show that the result of Algorithm \ref{alg:Graph-dissection-algorithm}
is in general not unique given the same graph depending on the sequence
of dissecting the graph.
\begin{example}
\label{exa:Let-the-variables}Let $x_{1}$, $x_{2}$, $x_{3}$, $x_{4}$
and $x_{5}$ be stochastically mutually independent random variables.
However, only $x_{1}$, $x_{3}$ and $x_{5}$ are measured and $x_{2}$
and $x_{4}$ are not considered. We have the following dependencies
$x_{6}=x_{1}\cdot x_{2}$, $x_{7}=x_{2}\cdot x_{3}$, $x_{8}=x_{3}\cdot x_{4}$
and $x_{9}=x_{4}\cdot x_{5}$. The corresponding graph and two possible
ways to dissect it are given in Figure \ref{fig:Two-possible-ways}
where we always return a set of complete subgraphs.

\begin{figure}[H]
\center\includegraphics[scale=0.35]{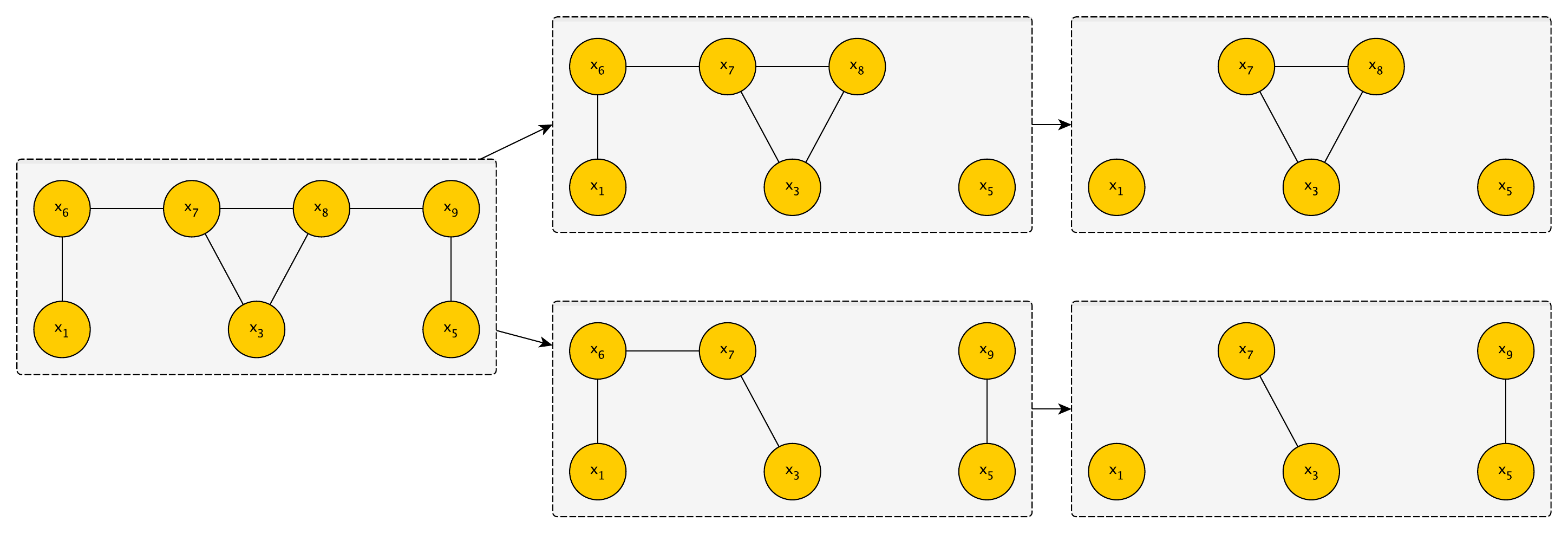}\caption{\label{fig:Two-possible-ways}Two possible ways to dissect a graph
with Algorithm \ref{alg:Graph-dissection-algorithm}. In the first
branch, we first remove $x_{9}$ and then $x_{6}$. In the second
branch, we first remove $x_{8}$ and then $x_{6}$.}

\end{figure}
\end{example}

The presented framework is not limited to the used techniques as explained
in the following.
\begin{rem}
\label{rem:The-dependency-graph}The dependency graph does not have
to be necessarily generated with a chi-square test, see Figure \ref{fig:Workflow-of-Section}.
Any test or measure that defines a dependency between two random variables
is suitable. For example, we can use the mutual information between
two variables \cite{papoulis2002probability,solomonoff2009information},
which was introduced by Shannon \cite[12.3.3.3]{billings2013nonlinear}.
In a subsequent step, the measure of dependency has to be transformed
into a statement if two variables are mutually independent. In the
case of the chi-square test, there is the level of significance to
evaluate if the hypothesis that both variables are mutually independent
have to be rejected. In the case of mutual information, analogously
a threshold has to be defined when we say that two variables are mutually
independent because they have not sufficient information in common.
If the measure is below that threshold, the variables are mutually
independent and mutually not independent if the value is above the
threshold. The threshold is necessary to filter for relations that
coincidently occur on a finite data set. Analogously, we can proceed
with the result of Algorithm \ref{alg:Graph-dissection-algorithm}.
Instead of a chi-square test to find out the variables that are linked
to our output variables, we can use the mutual information for example.
In addition, the mutual information between the random variables can
be used after a chi-square test between the variables from Algorithm
\ref{alg:Graph-dissection-algorithm} and the output variables. Taking
only variables whose mutual information with the output variables
is above a threshold can be used for model reduction via approximations.

In the next example, we demonstrate why it can sometimes be reasonable
to use the mutual information test after a chi-square test in order
to further reduce the model based on neglecting variables that only
contribute with small effects to a model.
\end{rem}

\begin{example}
\label{exa:Let-us-choose}Let us choose $x_{1}$ and $x_{2}$ as mutually
independent random variables with values between $0$ and $5$. We
define the output variable $y\coloneqq\begin{cases}
1 & \mathrm{\ if\ x_{1}+x_{2}\cdot10^{-0.5}\geq4}\\
0 & \mathrm{\ else}
\end{cases}$. The variables $x_{1}$ and $x_{2}$ are the basic variables to model
$y$ and the chi-square test will return both variables as linked
to the output variable since this test has binary results with respect
to the relation of a variable to an output variable. However, if the
mutual information test is performed, the score of $x_{1}$ is higher
of about one order of magnitude than the mutual information score
of $x_{2}$. Depending on the purpose and accuracy of modeling, a
model reduction to $x_{1}$ might be reasonable. 

Furthermore, if it holds that $x_{3}=x_{1}+x_{2}\cdot10^{-0.5}$,
then the output variable $y$ can be modeled by $x_{1}$ and $x_{2}$
but also just by $x_{3}$ if the values of $x_{3}$ are also given/measured.
Since Algorithm \ref{alg:Graph-dissection-algorithm} would always
return $x_{1}$ and $x_{2}$, the number of input variables could
be reduced without approximations by a detailed analysis of the dependencies
of the variables returned by Algorithm \ref{alg:Graph-dissection-algorithm}
(the set $X'$) and the other variables of the set $\tilde{X}=\left\{ x_{1},x_{2},x_{3}\right\} $.
This is always the case if an output variable depends just on the
projection of some variables. In our example, the projection is the
sum of two variables and in addition this projection is a further
measured variable in the system. Reducing the model by finding projections
of relevant variables is related to feature engineering where new
features are generated from original raw data by mathematical transformations
like multiply each two features to obtain the value of their product.

In the first case, information is neglected and the information of
the data set is approximated. In the second case the model reduction
is exact where no information is neglected but presented in a purposeful
representation.
\end{example}

In case that the dependency graph has many nodes, which means many
variables are considered, calculating the total adjacency matrix and
finding a set of minimal cardinality that dissects the graph can be
time consuming due to the quadratic scaling of the calculation of
the adjacency matrix and the polynomial complexity for finding a minimal
cut, see \cite{esfahanian2013connectivity}. Consequently, in order
to accelerate the calculations, Algorithm \ref{alg:Graph-dissection-algorithm}
can be applied to a subgraph of the dependency graph. Thus, only the
adjacency matrix for this subgraph has to be calculated first in which
the results of the chi-square tests for the corresponding variables
are stored for later calculations. The total dependency graph is processed
in parts where variables of each subgraph are removed which form a
typical structure for random variables that are a function of other
variables. Once no node in any subgraph cannot be removed anymore,
we apply Algorithm \ref{alg:Graph-dissection-algorithm} to the remaining
total graph. The adjacency matrix of the total graph defined in (\ref{eq:adja ma})
may not be complete after this procedure since chi-square tests of
removed variables with remaining variables are saved. However, if
we need some entries for further analysis, we can purposefully calculate
them. The advantage is that we reduce the chi-square tests for variables
removed within a subgraph compared to the case where we calculate
the total adjacency matrix for all the nodes. In addition, sets of
minimal cardinality that dissect a graph can be found faster in smaller
graphs due to the polynomial scaling of the corresponding algorithm.
We summarize the procedure in Algorithm \ref{alg:PVA Algo} and call
this algorithm the PFA. We call the result of Algorithm \ref{alg:PVA Algo}
the principal features.

\begin{algorithm}[H]
\begin{enumerate}
\item Choose the maximal number of nodes $n_{s}\in\mathbb{N}$ per subgraph 
\item Generate disjunct sets of at most $n_{s}$ nodes that cover the total
set of nodes
\item Generate the corresponding entries of the adjacency matrix defined
in (\ref{eq:adja ma}) for these subgraphs if it is not already calculated
in previous calculations
\item Apply Algorithm \ref{alg:Graph-dissection-algorithm} to each subgraph
\item If no node has been removed from any subgraph: Consider the total
graph of the remaining nodes and do Step 3 and Step 4, then Stop
\item Generate disjunct sets of at most $n_{s}$ nodes from the remaining
nodes and go to Step 3
\end{enumerate}
\caption{\label{alg:PVA Algo}PFA algorithm}
\end{algorithm}

In our implementation of Algorithm \ref{alg:PVA Algo}, all the remaining
nodes after Step 3 are sorted in ascending order and packed one after
another into sublists with $n_{s}$ elements starting with the first
node where the last sublist has at most $n_{s}$ elements if the number
of nodes is not an integer multiple of $n_{s}$. Since Algorithm \ref{alg:PVA Algo}
is iteratively applied to subgraphs of the total dependency graph,
Lemma \ref{lem:Algorithm--is} holds analogously for Algorithm \ref{alg:PVA Algo}.
However, in the view of Example \ref{exa:Let-the-variables}, the
choice of the subgraphs may influence the final result, i.e. the combination
of principle features.

The complexity of Algorithm \ref{alg:PVA Algo} scales as follows.
As mentioned before, generating the adjacency matrix scales quadratically
and finding a minimal cut of the graph scales polynomially regarding
the number of the nodes. However, by performing the calculations for
the adjacency matrix just for the necessary nodes and applying the
min cut algorithm to subgraphs reduces a lot of calculation effort.
How much this reduction of calculations decreases the runtime depends
on the particular topology of the dependency graph. 

\section{An evaluation of the PFA based on error identifications of a data
center\label{sec:An-application-of} }

In this section, we demonstrate how the PFA can be use to identify
relevant variables of a data set. The data in this section is collected
from a data center environment. The variables in this context describe
parameters of servers and are called metrics. From this metrics the
state of a server can be inferred. The information about the state
can be for example used in self healing systems that automatically
correct issues and thus enhance the reliability of cloud applications
\cite{chen2019outage,levy2020predictive}. For an efficient implementation
of a self healing system, it is important to focus on the metrics
containing purposeful information. 

First, we describe the implementation of the PFA and subsequently
show results. The implementation consists of how the data set was
acquired and what programming libraries are used. The presented framework
is implemented in Python 3.7. The used Python functions are specified
later. In the subsequent evaluation part, the results of the PFA are
compared to different strategies to reduce the data set, like the
principle component analysis (PCA) \cite{jolliffe2002principal} and
a minimize redundancy maximize relevance (mRMR) method \cite{peng2005feature}.
Furthermore, the post-processing of the results of the PFA with the
mutual information is demonstrated and the prediction accuracy of
different models trained on the sets of input variables are presented.

The data set consists of 2154 metrics that represent different parameters
of a server, see \newline https://learn.netdata.cloud/docs/agent/collectors/collectors
for a documentation of the parameters. A measurement of all the metrics
at the same time point is called a data point. 

Next, the acquisition of the data set is explained. The dataset used
in this section was generated using 15 physical servers with identical
hardware. We used a fault injection to transfer those 15 servers into
a faulty state. The failures considered affect the central processing
unit (CPU) and the Random Access Memory (RAM). The failure are caused
by scripts that are executed on the servers. The scripts use a tool
called stress (https://linux.die.net/man/1/stress) which acts as a
faulty program to manipulate the percentage of CPU and used RAM. The
fault injection scripts, which follow a certain structure, are shown
in Algorithm \ref{alg:Fault-Injection-Script}.

\begin{algorithm}[H]
\begin{enumerate}
\item Initialize all 15 servers
\item Start monitoring
\item Wait some time
\item Inject fault
\end{enumerate}
\caption{\label{alg:Fault-Injection-Script}Fault Injection Script}
\end{algorithm}

Initializing all 15 servers means that the script checks if it can
establish a connection to all nodes in the system and if so it sets
up OpenStack (https://www.openstack.org). OpenStack is a Cloud Operating
System with processes for managing different parts of the servers'
hardware. Afterwards the script uploads the required shell scripts
which cause the failures later on. After successfully doing so monitoring
is initiated. With the monitoring tool the measurements of the servers'
parameters are done to generate the data points. For monitoring, we
used a tool called Netdata (https://www.netdata.cloud). Netdata is
an open-source-tool that allows us to monitor and store a server's
state in real-time. The metrics were requested and stored for each
of the 15 servers each second. One failure scenario lasts 500 seconds
to have sufficient time to collect data from the server in a non-faulty
and a faulty state. Thus, a measurement series provides $15\cdot500=7500$
data points for each case where each data point consists of 2154 values.
At the start of monitoring the script randomly waits between 100 and
400 seconds to let the system run in a non-faulty state for some time.
This variation of the starting point for the fault injection results
in a balance between faulty and non-faulty data points. After that
time the failures are initiated remotely on all 15 servers at the
same time to parallelize the data acquisition. Each data point is
classified with 1 if the server was in an error state and 0 if the
state of the server was error free. The starting point of the fault
injection is also the point where the labels in the dataset switch
from 0 (non-faulty) to 1 (faulty). Even though all 15 servers receive
the same treatment, the measured results vary due to noise caused
by the OpenStack processes running on them. Thereby every server provides
slightly different data points. Several shell scripts able to cause
CPU and RAM failures have been uploaded to the nodes. For the CPU
failures, the stress tool claims a certain percentage of available
processes to force the desired workload. For RAM, the stress tool
allocates the desired percentage of available megabyte in memory.
As Netdata not only monitors the general CPU and RAM usage but also
the usage per application and user, not only two but multiple, possibly
redundant, metrics change at once during fault injection. 

As a measure of how many information a selection of metrics has, we
use different machine learning models and rate their prediction accuracy.
We use the fact that for a sufficient prediction accuracy it is necessary
that the used metrics carry sufficient information regarding the prediction.
We are aware of the fact that the reversal does not hold in general.
A reason could be that a selection of metrics might contain all the
information of a data set, however the chosen machine learning model
is not a good choice for constructing the functions to calculate the
relation between input metrics and output features, i.e. the prediction
values. The machine learning models act as a quick test, like an oracle,
answering the question if from a selection of features a sufficient
prediction accuracy can be done once the model is trained on the features
from the PFA.

Since the intention of the work is rather focussing on the preprocessing
of data than using the processed data with various models, like machine
learning models, we use the Python framework sklearn for an easy implementation
of classifiers based on neuronal networks (NN) and support vector
machines (SVM). The NN is implemented with the mlpClassifier from
Python sklearn.neural\_network and the SVM is implemented with SVC
from Python svm.

In order to evaluate the prediction accuracy the r2-accuracy score
(sklearn.metrics accuracy\_score) is used as well as the number of
wrongly classified data points from the confusion matrix.

Since the training of the NN is based on stochastic optimization methods,
we perform any training and the subsequent prediction based on a feature
selection 100 times. The standard deviation of the r2-accuracy score
and of the wrongly classified data points is not zero. This shows
the influence of the stochastic optimization routine that is used
to train the NN. A further reason for performing training and prediction
100 times is that in some experiments features are randomly chosen.

The PCA is implemented with the PCA from sklearn.decomposition. A
notion of the basic concept of the PCA can be found in the Related
methods section.

The mRMR method is implemented with the C++ source code from http://home.penglab.com/proj/mRMR/.
The basic description of the mRMR method can be found in the Related
methods section. To use the mRMR method as implemented in the C++
code, a discretization of the continuous variables is necessary. The
discretization in the C++ code is performed as follows. The boundaries
for the bins are set by the mean $\pm$ multiples of the standard
deviation (std) of the data set. We choose 1 as the maximum of a multiple
of std and obtain sufficient results of prediction accuracy. Consequently,
according to the description of the source code the boundaries of
the bins are mean $\pm$ $\left\{ 0,0.5,1\right\} \cdot$std. When
we execute the code we allocate memory for 25000 data points and 2500
variables. We use the implementation in the mutual information difference
(MID) mode. This mode selects the set of metrics such that the difference
of the sum of all mutual information between the selected metrics
and the output function and the sum of all mutual information among
the selected metrics with each other is maximized.

Before we use the NN, SVM or PCA, the data is normalized with the
MinMaxScaler from sklearn.preprocessing. This scaler is fitted on
the training set and applied to the train and test set. 

The chi-square test is implemented in the Python function chisquare
from the Python package scipy.stats.

The data set is randomly split into a train and a test data set where
we have $80$\% of the 30000 data points in the train data set. The
remaining data points are put into the test data set. The PFA is performed
on the training data set if not otherwise stated.

In this section, Algorithm \ref{alg:PVA Algo} is used. In order to
dissect the graphs with Algorithm \ref{alg:Graph-dissection-algorithm},
the Python routine minimum\_node\_cut from the Python package networkx
is used. The minimum\_node\_cut function is a flow based algorithm
to generate a set of nodes of minimal cardinality that dissects the
corresponding graph upon removing this set of nodes. For details,
see the documentation of networkx or \cite[Algorithm 11]{esfahanian2013connectivity}.

The result of the PFA is further processed with a chi-square test
to obtain these metrics on which the function which labels the data
points depends on. The function that labels the data points if a data
point corresponds to a faulty or a non-faulty state is the output
function. Metrics that belong to a node of a subgraph consisting of
more than one node are proceeded as follows. If the output function
is not independent of one metric corresponding to the considered complete
subgraph, we choose all the metrics corresponding to the current complete
subgraph for the selection of the relevant input metrics. For the
reasoning, see the paragraph starting on page \pageref{exa:We-see-that}. 

For further model reduction, we determine the mutual information of
each metric with the output function and take only the metrics above
a certain threshold $\theta>0$. The threshold $\theta$ is specified
for each experiment, see e.g. the experiment whose results are presented
in Table \ref{tab:Number-of-metrics}.

Next, it is explained how we use the PFA for the analysis of the data
set introduced above. For the binning, i.e. the discretization of
the co-domain of the random variables, we use Algorithm \ref{alg:Discretize-the-co-domain}
with $\nu=500$. By the choice $\nu=500$, the chi-square test had
at least 5 data points for each expected frequency of the joint outcome
of two metrics in any calculation, see Remark \ref{rem:We-remark-that}
for details. For any chi-square test in this section the hypothesis
that two variables are independent is rejected on a level of significance
of $1\%$. This level of significance is a common choice and provides
reasonable results as well in the following of this section. Next,
Algorithm \ref{alg:PVA Algo} is used with $n_{s}=50$. We experienced
that the subgraphs in this section consisting of at most 50 nodes
can be processed in reasonable time from the minimal\_node\_cut function.

The experiments are performed on a laptop with a 2.3 GHz 8-Core Intel
Core i9 processor with 32 GB 2667 MHz DDR4 memory. The time needed
for one run of the PFA with the settings discussed in the previous
paragraph on this laptop takes about 4 minutes.

Our first experiment is performed as follows. On the train data set
the PFA extracts 206 metrics. In order to compare the prediction accuracy
of the NN and the SVM with regard to different selections of metrics
the mean values of the r2-accuracy score and the mean number (\#)
of wrongly predicted data points are given in Table \ref{tab:Statistics-for-different}
with the corresponding standard deviations (std) from 100 training
and subsequent prediction sweeps. The results of the NN and the SVM
based on these 206 metrics after the training on the train data set
and the subsequent prediction on the test set can be seen in Table
\ref{tab:Statistics-for-different}, first and second column. In comparison,
the NN and the SVM are trained on randomly chosen 206 metrics that
are non-constant, third and forth column. There are 1536 non-constant
metrics in the data set. Non-constant means that there is a measurement
of this metric which is different from the other ones. In the fifth
and sixth column of Table \ref{tab:Statistics-for-different}, there
are the results where the NN and the SVM are trained on all the non-constant
metrics. 

\begin{table}[H]
\center%
\begin{tabular}{|c|c|c|c|c|c|c|}
\hline 
 & PFA NN & PFA SVM & random NN  & random SVM & all NN & all SVM\tabularnewline
\hline 
\hline 
r2-accuracy mean & 0.9973 & 0.9982 & 0.9875 & 0.9580 & 0.9932 & 0.9663\tabularnewline
\hline 
r2-accuracy std & 0.0029 & $10^{-16}$ & 0.0179 & 0.0207 & 0.0102 & 0.0\tabularnewline
\hline 
\# wrongly classified mean & 16.2 & 11.0 & 75.3 & 251.97 & 40.9 & 202.0\tabularnewline
\hline 
\# wrongly classified std & 17.7 & 0.0 & 107.7 & 124.4 & 61.2 & 0.0\tabularnewline
\hline 
time of training mean & 8.4584 & 3.9269 & 12.7301 & 5.2470 & 32.5983 & 25.4406\tabularnewline
\hline 
time of training std & 1.7704 & 0.2962 & 3.7647 & 2.6496 & 3.5409 & 0.6769\tabularnewline
\hline 
\end{tabular}\caption{\label{tab:Statistics-for-different}Statistics for different methods
of the selection of the input metrics for two different models. The
first model is a neuronal network (NN) classifier implemented in mlpClassifier
of sklearn. The second model is a support vector machine (SVM) classifier
implemented in SVM of sklearn. In the column PFA NN and PFA SVM there
are 206 metrics selected with the PFA on which the training is performed.
In the column random NN and random SVM there are 206 randomly chosen
metrics from all the non-constant metrics of the data set on which
the training is performed. In the columns all NN and all SVM, the
models are trained on all the non-constant 1536 metrics of the data
set on which the training is performed. In the last two rows there
are the mean and std time in seconds to train the corresponding model
on the chosen metrics. }

\end{table}

\label{The-results-of}The results of Table \ref{tab:Statistics-for-different}
can be interpreted as follows. In the case where the metrics are selected
from the PFA, the NN and the SVM perform equally well. In the case
of fixed metrics and the classification by the NN, i.e. the columns
named with PFA NN and all NN, the std is greater than zero. This comes
from the stochastic training method. We assume that the reason why
the performance on the PFA metrics is better than on all the non-constant
metrics is that small models are easier to train. Since the information
is concentrated in models using just the relevant metrics, the method
can easier distinguish if a small progress in learning is due to an
already optimal trained model or if only few weights of the model
change since only a little number of metrics is linked to the output
variable and thus only these linked metrics change during learning
anyway. Illustratively spoken, the information from the PFA is concentrated
and not diluted. In comparison to the case where the 206 metrics are
chosen randomly, we see that the metrics chosen with the PFA framework
perform much better. Summarizing Table \ref{tab:Statistics-for-different},
the learning on the PFA selected metrics works better and faster than
on all the other two cases independent of the considered machine learning
model.

There are high accuracy scores for the NN and the SVM, see Table \ref{tab:Statistics-for-different}
in any cases. Consequently, we are confident that the corresponding
prediction function can be well approximated by the chosen models.
We summarize that the PFA provides metrics that contain sufficient
information from the data set for the prediction if a server state
is faulty or non-faulty. 

Next, an NN and an SVM is trained each 100 times on a set that is
transformed with the PCA providing 206 components carrying 99.7252
\% of the variance of the original data set. The NN has an r2-accuracy
mean of 0.9983 with std 0.0002 and 10.25 wrongly classified data points
with std 1.0805. The needed time for a training is in mean 3.3380
seconds with std 0.1645 seconds. The SVM has a mean r2-accuracy score
of 0.9674 with 0.0001 std and 195.49 wrongly classified data points
with 0.6707 std. Since the NN prediction accuracy based on the PCA
is comparable with the prediction accuracy of the NN based on the
PFA, the data set transformed with the PCA contains the necessary
information for learning the function for the prediction if the server
is in a faulty or a non-faulty state. However, there is a difference
in preprocessing data sets with the PCA and the PFA. The PCA determines
a linear combinations of the original features to transform the data
set into a new data set where the new features are the linear combinations
of the original features. The PFA extracts the features from the original
data set that carry the relevant information. In the presented example
the SVM cannot learn from data set transformed with the PCA as good
as on the original features presented from the PFA. From the point
of training a machine learning model the PCA and the PFA might be
considered equally. From the point of modeling and interpretability,
the PFA has the advantage of providing a set of input features of
the original set from where the modeling can start. To start from
the original features can enhance the interpretability of the model.
The PCA linearly transforms the original features into a new data
set. For this transformation all the original features may be needed.
In case of the PFA, all the features that are not returned as relevant
may be neglected and do not need to be measured when applying the
model in a use case. Furthermore, learning on the original data set
may provide advantages for explaining the models.

Since from Table \ref{tab:Statistics-for-different} we can extract
that the NN and the SVM perform equally well on the PFA metrics and
the NN performs better in the non PFA cases, the remaining part of
this section is carried out with the NN if not otherwise stated.

Next, the mutual information of the 206 metrics from the PFA and the
output variable (function that labels if a data point is from a faulty
or a non-faulty server state) are determined. Then only the metrics
that are above a certain threshold $\theta$ are chosen. We take $\theta\in\left\{ 0.05,0.1,0.15,0.2\right\} $
as a reasonable selection where the effect of approximating information
can be seen quite well. Analogously, each experiment is performed
100 times and the resulting statistics of the experiments are presented
in Table \ref{tab:Number-of-metrics}. If the metrics are randomly
chosen, new metrics are selected after each sweep.

\begin{table}[H]
\center%
\begin{tabular}{|c|c|c|c|c|}
\hline 
 & r2-accuracy mean & r2-accuracy std & \# wrongly classified mean  & \# wrongly classified std\tabularnewline
\hline 
\hline 
PFA 0.05 140 & 0.9981 & 0.0003 & 10.9 & 1.8\tabularnewline
\hline 
rand 140 & 0.9813 & 0.0210 & 112.3 & 126.0\tabularnewline
\hline 
PFA 0.1 122 & 0.9979 & 0.0020 & 12.4 & 11.8\tabularnewline
\hline 
rand 122 & 0.9772 & 0.0327 & 136.6 & 196.4\tabularnewline
\hline 
PFA 0.15 114 & 0.9973 & 0.0003 & 16.2 & 2.1\tabularnewline
\hline 
rand 114 & 0.9772 & 0.0252 & 136.6 & 151.3\tabularnewline
\hline 
PFA 0.2 107 & 0.9189 & 0.0018 & 486.4 & 11.1\tabularnewline
\hline 
rand 107 & 0.9774 & 0.0281 & 135.6 & 168.3\tabularnewline
\hline 
\end{tabular}\caption{\label{tab:Number-of-metrics}Number of metrics that are further reduced
by taking only the metrics whose mutual information with respect to
the output variable are above a threshold $\theta$. In the row starting
with PFA, we have the results of the selection of variables after
the PFA and $\theta$ equalling the number following PFA. The number
that follows is the number of metrics from the PFA that are above
the corresponding threshold and are used for a prediction. In the
subsequent row starting with rand the experiment is performed on randomly
chosen metrics where the number of used metrics equals the subsequent
number.}

\end{table}

In Table \ref{tab:Number-of-metrics}, the prediction accuracy of
the selection of variables whose mutual information with respect to
the output variable is above $\theta=0.05$ increases compared to
the model learned from all the 206 variables from the PFA of Table
\ref{tab:Statistics-for-different}. A reason might be that the smaller
model can be trained easier and that this advantages predominates
the fact that information is deleted from the data set. A further
reason can be that variables are removed that might be selected by
the PFA due to spurious correlations which coincidently may exist
in our data set. If we proceed, the prediction accuracy decreases
since more and more metrics that are relevant for the prediction are
neglected, see the column r2-accuracy mean of Table \ref{tab:Number-of-metrics}.
Since the PFA selects metrics such that the redundancy is removed,
we cannot construct information once deleted by the other metrics.
For the case of $\theta=0.2$ much relevant information is deleted
since the randomly chosen metrics perform better. Further the relative
small std of wrongly classified data points may be an indicator that
the misclassification is due to a systematic loss of information that
is caused by the successively reduction of number of metrics. 

In order to demonstrate that focusing on the metrics with the highest
mutual information with respect to the output variable, does not necessarily
provide the best results, we perform the following experiment. We
take the mutual information of all metrics with respect to the output
variable and choose the metrics whose mutual information is above
the threshold $\theta=0.7752$. This threshold provides 208 metrics.
We perform the training of the NN on these 208 metrics 100 times and
obtain a mean r2-accuracy score of 0.9102 with a standard deviation
of 0.0164 and a mean of 510 with std of 98 wrongly classified data
points. Again there is a small std. This may indicate that information
is systematically missed. The reason is that there is now a high redundancy
within this set of input variables but this set does not cover the
total information necessary for a sufficient classification. The result
is that just taking variables that have a high mutual information
score is not sufficient for a good prediction since the input information
is not necessarily independent and thus may contain redundant information
instead of the total information of the data set. This demonstrates
that also many metrics with each only a small contribution of mutual
information with respect to the output variable can make a valuable
contribution to a correct prediction. An illustration is a function
that depends on many variables. If each variable has the same mutual
information score with the function, the score of each variable decreases
the more of such variables the function depends on. By just focussing
on a threshold of mutual information, many variables of one function
can be deleted. Consequently for a model reduction it is not always
the best to focus just on metrics with the highest mutual information
since the interplay of metrics with only a little mutual information
each with the output function can be important. The discussion of
this paragraph shows that reducing the variables with the PFA is a
good way to end up with metrics being relevant for a prediction since
the prediction accuracy of Table \ref{tab:Number-of-metrics} is much
better than by choosing the metrics with a single mutual information
test.

Next, the PFA is compared with the mRMR method on the data set that
is used for the experiments presented in Table \ref{tab:Statistics-for-different}.
For the purpose of comparison a different number of features is extracted
by the mRMR method and a classification with an NN and an SVM is performed
on the same data set as used for the experiment where the corresponding
results are presented in Table \ref{tab:Statistics-for-different}.
The training and subsequent prediction is performed on each set 100
times. The results are presented in Table \ref{tab:Results-from-the}.
The fact that the intersection of the 206 PFA metrics with the 206
mRMR method metrics consists of only 22 common metrics shows that
both methods work differently. The aim of the PFA is to find the arguments
in data sets with which a considered function can be modeled. The
aim of the mRMR method is to find a set of metrics that provides the
best prediction accuracy where the different metrics in this set are
supposed to be as mutually independent of each other as possible.
The mRMR method does not necessarily distinguish between features
that are functions of others and the corresponding arguments. If an
output function depends on projections of features rather than on
the metrics explicitly, the mRMR method is likely to choose the features
that represent the projection. For example in a data center the CPU
utilization depends on the sum of all the CPU utilization of any process
that is running in the data center. If an output function just depends
on the sum of the CPU utilization the mRMR method is likely to pick
the metric describing the total CPU utilization instead of the metrics
describing the single CPU utilization of any process running in the
data center which is more likely using the PFA. In Table \ref{tab:Results-from-the},
we see that the prediction accuracy sharply improves when taking more
than 175 metrics. The results in Table \ref{tab:Statistics-for-different},
Table \ref{tab:Number-of-metrics} and Table \ref{tab:Results-from-the}
show that both methods work successfully. Both methods can provide
useful sets of metrics from where a modeling can start. We remark
that the number of 206 metrics obtained from the PFA and possibly
needed to describe the label function well in this case is a well
working initial guess for the number of metrics the mRMR method is
supposed to return. To obtain the 206 metrics from the used implementation
of the mRMR method, it took about 2.5 hours and about 5 hours to obtain
300 features.

The prediction accuracy in Table \ref{tab:Results-from-the} improves
beyond 206 metrics. This improvement can be explained as follows.
As discussed e.g. after Remark \ref{rem:If-we-choose} removing a
feature that is a function of other features can cause a loss of information
if not all arguments the function depends on are contained in the
considered data set. The mRMR method includes step by step any feature
that shares mutual information with the output function and that has
as little mutual information with the already selected features as
possible. It can happen, if the number of features to be selected
is sufficiently high, that also features being functions of features
are selected in the cases where arguments of the function are missing
to increase the amount of information in the set of selected features.
However if the selected features exceed a threshold, the gain of further
information by the additional features is small and wasted by computational
issues as discussed in the paragraph starting on page \pageref{The-results-of}.
The inclusion of features that are functions may be desired if the
focus is on compressing a data set to obtain the best prediction possible.
However, in the case where the focus is to analyze the relations within
a data set, having features that are functions in the set of independent
arguments may be undesirable since the aim is to find basic features
with which all functions can be constructed to understand the relations
and dynamics how the dependent features evolve given the input variables.
In this case, it can be helpful that we are made aware by a not sufficient
prediction accuracy that the data set is missing some further relevant
features that should be included.

\begin{table}
\center%
\begin{tabular}{|c|c|c|c|c|c|c|}
\hline 
 & 150 & 175 & 206 & 230 & 300 & 350\tabularnewline
\hline 
\hline 
NN mean r2-accuracy & 0.9426 & 0.9987 & 0.9990 & 0.9990 & 0.9958 & 0.9905\tabularnewline
\hline 
NN mean \# wrongly classified & 344.18 & 7.77 & 6.17 & 6.11 & 24.73 & 56.75\tabularnewline
\hline 
NN mean training time (s) & 12.27 & 9.74 & 10.32 & 7.72 & 11.27 & 12.07\tabularnewline
\hline 
SVM mean r2-accuracy & 0.9422 & 0.9985 & 0.9985 & 0.9985 & 0.9988 & 0.9663\tabularnewline
\hline 
SVM mean \# wrongly classified & 347.0 & 9.0 & 9.0 & 9.0 & 7.0 & 202.0\tabularnewline
\hline 
SVM mean training time (s) & 1.86 & 1.77 & 2.37 & 2.27 & 3.22 & 3.51\tabularnewline
\hline 
\end{tabular}

\caption{\label{tab:Results-from-the}Results from the mRMR method for different
number of metrics to be selected denoted in the columns. In the rows
the mean classification accuracy of the NN and the SVM are presented.
Furthermore, the mean training time in seconds (s) for the NN and
the SVM is given.}
\end{table}

In the following, we discuss how to make the PFA robust with respect
to the data set's statistic properties upon interchanging parts of
the data, i.e. by making new train data sets. For the next experiment,
the train data set is reduced to 95\% of the data. The data points
to be left out are chosen randomly. The sampling of new data sets
with 95\% of the original training set is performed 5 times. The sets
of the metrics returned by the PFA on each data set are intersected.
Subsequently, 161 common metrics are obtained. The training and the
prediction of the NN is repeated analogously as described above 100
times on the original train and test data set. We obtain the following
results. The r2-accuracy mean is 0.9984 and the std equals 0.0022.
The mean of wrongly classified data points is 9.71 and the std is
13.1044. If the 182 metrics are used that we obtain from the PFA after
applying it to the total data set (joining train and test data), the
corresponding r2-accuracy score is in mean 0.9945 and the std is 0.0036.
Th mean of wrongly classified data points is 33.1 and the std is 21.8.
We interpret this result as follows. Strong dependencies and the corresponding
stochastic properties are robust with respect to interchanging data
points. However, relations that are only coincidently in a data set
are not robust with respect to interchanging data points. As experienced
in this case interchanging data points can significantly change the
stochastic properties of some parts of the data. On the other hand,
there are relations that are robust with respect to interchanging
data. This relations are extracted with the PFA and a sufficient prediction
is possible as the results indicate. Consequently, by using the intersection
of the sets resulting from the application of the PFA to different
splittings of a data set extracts the strongest dependencies that
are likely to be non-spurious relations between the corresponding
metrics. Hence, the signal-to-noise-ratio is improved enhancing building
significant models. In the case that instead of the PFA the PCA is
used on the train data set, used for the experiment presented Table
\ref{tab:Statistics-for-different}, to extract 161 components carrying
99.5679 \% of the variance of the data, we obtain a mean r2-accuracy
score of 0.9983 with 0.0002 std and a mean of 10.24 wrongly classified
data points with std 0.9287. The mRMR method provides a mean r2-score
of 0.9986 with 0.0003 std and 8.61 mean wrongly classified data points
with 1.9692 performing the training and the prediction 100 times on
the train and test data set as used for the experiment presented Table
\ref{tab:Statistics-for-different}. The variations of the results
of the mRMR method regarding performing it on different slightly varied
data sets (90\% randomly selected from the train data set) is small.
Only the positions of the single metrics in the returned lists slightly
differ sometimes. Consequently, the intersection of these lists contains
almost all the metrics of the results up to 2 metrics considering
lists of 200 metrics from 3 sweeps.

A related experiment to the experiment from the last paragraph ist
the following. The samples in Step 2 of Algorithm \ref{alg:PVA Algo}
can be sampled randomly instead of generating the sub sets of nodes
by dividing the ascending sorted list of nodes by packing $n_{s}$
nodes one after another into a list. For each experiment with randomly
sampled sub sets in Algorithm \ref{alg:PVA Algo}, the results of
the PFA slightly differ in our three runs of the PFA. The number of
selected metrics is between 192 and 199. Possible reasons for the
slight variation are discussed on page \pageref{exa:We-see-that}
ff. The sets of metrics, including the set that we use for the experiments
above with the 206 metrics, are the same up to about 20 variables.
We train an NN and an SVM model on each set 100 times and obtain the
following mean values. The r2-accuracy score for the NN model ranges
between 0.9929 and 0.9988 with std 0.0013 and 0.0076. The wrongly
classified data points range between 7.27 and 42.55 with std 7.5404
and 45.7957. For the SVM, the r2-accuracy score equals 0.9985 and
the wrongly classified data points are 9.0 in any case. If an SVM
model is used, the results on the used data set indicate that the
SVM model is not sensitive to small variations of the result of the
PFA.

We can summarize the last tow paragraphs as follows. It can be a possibility
to perform the PFA on different randomly sampled train data sets and
let the samples in Step 2 of Algorithm \ref{alg:PVA Algo} be selected
randomly. On any result of the PFA machine learning models can be
trained and we can choose the set of input metrics with the best prediction
accuracy. This result can be compared to the result that is obtained
when sampling randomly different train data sets, perform the PFA
on each data set and intersect the results. 

In the next experiment, we investigate the PFA on different splittings
of our total data set starting with 80\% of the total data set and
decrease the part of data points that is interchanged in each experiment
to test if there is a threshold when the stochastic properties of
the data set are robust with respect to interchanging data. For this
experiment, we join the training and the test set and split the total
data set in randomly chosen train and test sets with a certain percentage
of data points for the train data set. We repeat the procedure five
times with the same ratio of elements in the train and test data set.
The metrics that are obtained are compared if in each of the five
sweeps the result of the PFA is the same. This is the case if we only
interchange $0.1\%$ of the total data set. We remark that this robustness
how many percentage of the data points can be interchanged until the
results of the PFA change is a property of the data set.

The framework can module wise be applied to analyze relations of coupled
systems as described in the following remark.
\begin{rem}
Once the metrics are identified that are related to model an output
function, e.g. a function that labels measurements, we can interpret
the identified input metrics in turn as functions that are influenced
by external processes that are not contained in the current data set.
An example can be processes running on a server. Then corresponding
measurements can be performed resulting in a data set containing the
values for the new output functions and processes running. This set
is analyzed with the PFA to obtain the new input variables, e.g. the
processes that are related to the new metric output variables. Iteratively,
we can build a model for any subsystem. Thus we can glue these well
validated submodels together via the corresponding input and output
functions to built a total model module wise. For a data center example
it means that once the internal processes are understood, it can be
investigated what external processes influences the internal processes.
Thus further explanations can be found on how an error of a server
may be caused by a process. Of course, if we are just interested which
processes are involved in error causing events, the intermediate step
via the metrics is not needed. However, this demonstrates how the
presented framework can be used to analyze the relations in a complex
system step by step and how well validated submodels can be recycled
to answer future research questions instead of starting for any issue
with a new model from the scratch.
\end{rem}

\section{Application of the PFA to a single cell data set\label{sec:Application-of-the}}

This section is intended to show the usability of the PFA in biology.
In particular, to analyze gene expression profiles of different cell
types. The features in this scenario are the expression levels of
the different genes. With the opportunity to measure the expression
patterns of single cells, each of such a cell measurement is a data
point that provides the values for the features. On each of a data
point a corresponding output function can be defined, like a label
for the corresponding cell type from which the data point is measured.
Such cell types can be a tumor cell or a physiological cell of tissue.
Further, a stage of a stem cell in its maturation process and a corresponding
differentiated cell type. Once an output function is defined on the
data points, the PFA may return a set of genes that carries the information
to construct this function. Thus two different cell types could be
distinguished from their expression levels of the relevant genes.
Starting from these relevant genes, we can build a model to analyze
what alterations of gene expressions are responsible to transform
one cell type into another for example. 

The data set from the presented example contains different expression
patterns of macrophages from mice. The GEO data set with the accession
number of GSE134420 were downloaded under \newline https://pubmed.ncbi.nlm.nih.gov/31391580/,
among them one sample SIA0 (accession number: GSM3946323) was extracted.
The file begins with the barcodes of each cell and contains the ribonucleic
acid (RNA) count of each cell and gene. The RNA count is the number
of RNA molecules that are transcribed from a gene corresponding to
the expression level of each gene.

The data was carefully processed using Seurat package \cite{butler2018integrating,stuart2019integrative}
(version 3.2.0; https://satijalab.org/seurat) as follows. The low
quality cells, including broken cells with unusual high percentage
of mitochondrial gene expression (>7.5) and low expressed gene features
(<200) were firstly excluded from the data set. After a global-scaling
normalization based on the top 2000 variable gene features, 15 principle
components were selected to calculate the distance and find the neighbors.
A resolution parameter of 0.5 was then applied to classify these cells
into 7 big clusters and 1 small cluster (n<50). Annotation using gene
markers, naming the 7 big clusters, resulted into three interstitial
macrophages, one lining macrophage, two precursor cell types and one
possible fibroblast type. The gene expression pattern normalized with
the transformation $x\mapsto\ln\left(x+1\right)$ was exported into
a matrix for further analysis. The normalized expression pattern of
each cell is labeled according to which cluster a cell belongs. 

Instead of labeling the measurements by clustering the measurements
with Seurat, the labels can also be directly measured if the cells
can be sorted according to certain properties, for example with a
fluorescence-activated cell sorting \cite{picot2012flow}. 

We choose the cluster with label 1 (CX3CR1, interstitial macrophage)
and the cluster with the label 3 (lining macrophage). The processed
data set that we use in this section has 9513 different genes and
121 data points. The 121 data points consist of the cells of the two
clusters where the number of cells that belong to a corresponding
cluster is almost equal.

The implementation of the PFA is the same as in Section \ref{sec:An-application-of}.
In the current section, it holds $\nu=25$ in Algorithm \ref{alg:Discretize-the-co-domain}.
With the choice of $\nu$ no expected frequency in the chi-square
test is below 5 data points.

For the training of the machine learning models, the data is split
randomly into a train and a test data set where the train data set
contains $80$\% of each class of the labels of the total data points.

The PFA analysis takes about 75 seconds and returns 9 genes. The NN
and the SVM are trained each 100 times and subsequently predict the
cell type of a data point of the test data set depending on these
9 genes. The results are presented in Table \ref{tab:Prediction-accuracy-based}.
The discussion of the results is analogous to the ones of Table \ref{tab:Statistics-for-different}.
The results in Table \ref{tab:Prediction-accuracy-based} demonstrate
that the PFA can also work on small data sets like the one used in
the presented case consisting of only 121 samples. Moreover, the results
show the benefit of a preprocessing and an analysis of the data set
before building models when comparing the prediction accuracy on all
and the 9 genes. We can summarize that in the presented example the
selection of the PFA contains sufficient information to model the
difference of the two cell clusters. 

The results from the PFA can also be validated from a biological perspective.
The 9 genes are denoted with Agfg1, Slc50a1, Msn, Supt5, Dhdds, Pih1d1,
H2-Ob, Rps14, Atic in the data set. These genes indicate the difference
between the interstitial macrophage and lining macrophage with regard
to a specific metabolism induced by Rps14, Slc50a1, Atic and Dhdds,
antigen processing (H2-Ob) and cell proliferation/migration regulated
by Msn. The PFA offers a rapid way to understand the new features
of novel-reported cell subtypes, for instance in this case, the epithelial-like
lining macrophage which forms an internal immunological barrier in
the synovial lining. Compared to regular differential gene expression
analysis, the PFA offers efficiently an augmenting view to understand
the differences regarding features between cell types.

\begin{table}
\center%
\begin{tabular}{|c|c|c|c|c|c|c|}
\hline 
 & PFA NN & PFA SVM & random NN & random SVM & all NN & all SVM\tabularnewline
\hline 
\hline 
r2-accuracy mean & 0.9248 & 0.9600 & 0.5112 & 0.52 & 0.6660 & 0.5200\tabularnewline
\hline 
\# wrongly classified mean & 1.88 & 1.0  & 12.22 & 11.98 & 8.35 & 12.0\tabularnewline
\hline 
\end{tabular}

\caption{\label{tab:Prediction-accuracy-based}Prediction accuracy based on
the result of the PFA (9 genes) for an NN and an SVM (column PFA NN
and PFA SVM). In the columns random NN and random SVM the accuracy
of the prediction of NN, resp., SVM is presented for randomly chosen
9 genes with a non-constant expression pattern which are 6972 genes
from the 9513 genes. In the column all NN and all SVM the prediction
accuracy is given where the NN and the SVM are trained on the total
data set where only the genes with a non-constant expression profile
are considered. In the rows, the mean of the r2-accuracy score and
the mean number (\#) of wrongly classified cells of the 100 training
and prediction cycles are given.}
\end{table}

We conclude this section how the PFA can be further used to analyze
cell behaviors and the causing relations for the behaviors module
wise.
\begin{rem}
The PFA can be used to identify genes or proteins from where an output
function, e.g. a label of cells, can be modeled. Subsequently these
identified genes or proteins can in turn be defined as output functions
that are supposed to be modeled by different input variables or features,
respectively. For example the expression pattern of these genes can
be measured as well as external stimuli in the environment of the
cell, like concentration of hormones or other signal molecules. Then
the PFA can be applied to find out the external stimuli that effect
the corresponding genes. By iterating the process of modeling former
input variables with related other submodels, we can build a total
model from submodels by glueing them together via defining input-output-interfaces.
The PFA framework can help to analyze the data and identify the relevant
parameters for each submodel. By breaking a system into subsystems,
we obtain building blocks of understandable subsystems revealing the
driving causes for a behavior of a system. Moreover, the well validated
submodels can be recycled for new research questions without starting
from the scratch. An example can be the communication of immune cells.
Once an immune cell type is modeled regarding what signal proteins
it secretes upon certain stimuli, a model of the total immune system
can be modeled from the models of several different immune cells.
\end{rem}

The next remark is an example where an output function represents
continuous values.
\begin{rem}
In biology, the t-distributed stochastic neighbor embedding (t-SNE),
which is an unsupervised machine learning method, is used to projected
single cells into a plane according to their expression pattern \cite{kobak2019art,zhou2020visualization}.
With this method the cells can be clustered and the clusters can be
visualized. In order to study what genes contain the relevant information
for the clustering of the t-SNE, the PFA can be used as follows. Each
cell that is projected into the plain can be characterized by a two
dimensional vector representing the coordinates of the cell in the
plane. This vector-function is defined as the output function. The
PFA is now used to find the genes that are the arguments of this function
that maps the coordinates of each cell into the plane. By focusing
on the relevant genes that are responsible to infer the position of
the cell in the plane, a model can be worked out that describes how
to influence which gene to reproduce the observed transformation of
the cells. Illustratively spoken, the PFA can be used to learn from
unsupervised methods and subsequently transform this information into
a mechanistic model.
\end{rem}

\section{\label{sec:Related-methods}Related methods}

In this section, we discuss the relation of the presented framework
to other already existing methods to preprocess data sets and extract
relevant variables from data sets or reduce a data model.

There are other methods to reduce the dimensionality of a data set
by compressing the information of the total data set. For example
the principle component analysis (PCA) \cite{jolliffe2002principal}
decomposes a data set into a linear combination of the principal axes.
However, it is challenging for the PCA to model non-linear relations
between the variables. In the linear case, we may obtain a transformation
of the data set that can be described with less variables. These variables
are linear combinations of other variables from the original data
set and it may be challenging to interpret these new compound variables. 

Furthermore, there are rank correlation methods like, Spearman's or
Kendall's test, see \cite{zwillinger1999crc,shevlyakov2016robust}.
These methods can be used instead of a chi-square test, see Figure
\ref{fig:Workflow-of-Section}, to evaluate if two variables are independent
of each other. A threshold defining independency needs to be defined,
as discussed in Remark \ref{rem:The-dependency-graph}. However, for
the rank correlation methods, it is challenging to detect non-monotonic
relations between variables. The application of e.g. a chi-square
test does not have limitations concerning functional relations of
variables. In cases where a rank correlation methods is better suited
to analyze the dependencies of the features, we can use the presented
framework to compress the information of independency from the rank
correlation methods as summarized in Figure \ref{fig:Workflow-of-Section}
where the chi-square test is replaced by a corresponding rank correlation
method. Subsequently, the correlation between the features can analogously
be modeled into a corresponding dependency graph by defining thresholds
for no correlation of two features and the workflow is as depicted
in Figure \ref{fig:Workflow-of-Section}.

If techniques from non-linear systems identification like the FROLS
algorithm, see \cite{billings2013nonlinear} for details, are applied,
we need to specify the terms of which the model is supposed to be
generated. These models can be polynomials of the variables for example.
On the one hand, these polynomials can result in an enormous growth
of parameters for all the monomials to capture all the combinations
of variables. On the other hand specifying a model is not necessary
if the model is supposed to be learned by a machine learning algorithm.
The fit of a polynomial model by e.g. regression can be applied after
the application of the PFA just for the identified relevant variables. 

A similar problem may be faced if we have to choose kernel functions
to project non-linear dependencies of random variables to linear relations
\cite{lopez2013randomized}. In this case, we have to make assumptions
on the considered data by choosing a kernel in contrast to the presented
method where dependencies are analyzed without a previous assumption
on the functional structure of the model. 

There is a technique from neuronal networks, called autoencoder \cite{pawar2020assessment},
that compresses the information of input variables to a smaller set
of variables. However, the autoencoder needs to be trained on the
total data set, where we might have the curse of dimensionality. The
PFA is a tool to prevent this situation. Moreover, if the autoencoder
is trained, it is challenging to understand how the neuronal network
compresses the information of the data set and what the meaning of
these compressed output nodes, which are the input nodes for the modeling,
is. In particular, it is also challenging how the total data set is
related to the output nodes of the autoencoder. Consequently, it may
be hard to find out, if or which variables of the total data set are
necessary for the prediction anyway and which can be neglected. 

Another method of feature selection is to select a subset of features
and train machine learning models on this selection. Finally, the
selection of features is chosen that provides the best accuracy of
prediction. This method of feature selection cannot be applied when
there is a large number of features due to the combinatorial effort.
The PFA returns features corresponding to complete subgraphs. Proceeding
like in the paragraph starting on page \pageref{In-order-to}, only
subgraphs are returned where there is at least one feature of which
the output function is not independent. Since the features corresponding
to different complete subgraph are independent and thus represent
information that is not redundant, the procedure of combing different
features can remarkably be simplified. We choose a subgraph and train
a machine learning model on any combination of the features corresponding
to the chosen subgraph while we take all the other features that do
not correspond to the chosen subgraph but to the other features returned
by the PFA framework. We choose the combination of features that provided
the best prediction accuracy. If two selections are comparable, then
we choose the one with the smaller number of features. Then we proceed
in the same way with the next complete subgraph that has more than
one node. The PFA structures and thus reduces the possible combinations
of features to only a few number such that a fine tuning of a feature
selection can be performed by trying any combination. A discussion
why the complete subgraphs with more than one node can correspond
to features that can be possibly sorted out is from page \pageref{exa:We-see-that}
to page \pageref{exa:Let-the-variables}.

A class of methods for constructing causal networks are causal inference
methods, see \cite{colombo2012learning}. This class of methods uses
the conditional independence of random variables and subsequently
relates this conditional independence information in a graph, namely
a directed acyclic graph (DAG). Going through this graph in the direction
of the edges, starting from the nodes that do not have an incoming
edge (input nodes), can be interpreted as giving explanations for
the subsequent nodes in the sense that one quantity is the cause for
alterations of the other quantity represented by the corresponding
node. The causal inference methods and the PFA can be combined as
follows. Since the main intention of the PFA is to identify functions
of features and thus compress the information of a data set to the
essential argument features, the application of the PFA is in particular
useful in large scale problems to reduce the feature set where there
are many features and possibly more than one output function. The
reduction to the main features may save a lot of combinatorial calculation
effort. One possible combination of both methods is that the PFA can
be used to provide the input nodes from which the DAG can be constructed
by a causal inference method. The PFA can thus provide a reasonable
procedure to give roots from where a reasoning of a causal inference
method can start. This may also be an opportunity to obtain some reasonable
initial nodes from where to start creating a DAG again if the output
from a causal inference method is challenging to interpret. A further
possibility to combine both methods is where we would like to find
the essential model to calculate/predict output variables and would
like to have some features in between the input and output variables.
This can be useful to better understand the function that maps input
variables to output variables. For the described application we proceed
as follows. Once we have the input variables by the PFA, all the other
features that are stochastically independent of the output variables
are sorted out. Then we can apply a causal inference method to construct
a DAG as an explanation going from the input to the output nodes.
This may save calculation time since the feature space can be considerably
reduced.

An analogously discussion as for the interference methods holds for
methods building association rules for a data set, see \cite[Chapter 6 and Chapter 7]{han2011data}.
Association rules describe conditions (values that features have)
that lead to the outcome of another feature with conditional probability.

We would like to remark that finding a causality may not always be
the focus of modeling. For instance, le the statement ``condition
1 and condition 2 is true'' be equivalent to ``condition 3 is true''.
Then we could say that condition 1 and condition 2 cause condition
3 since no other conditions influence condition 3. In the case that
condition 3 is a disease, we usually would like to know the causes.
However, since we have the data available for all the three conditions,
maybe for some use cases, it is also interesting to build a model
to infer condition 2 from condition 1 and condition 3, not as a causal
model but just exploit functional dependencies to infer a quantity
from others, e.g. to infer something about the causes given the effects.
An analytic example is where $x_{1}=2x_{2}$, $x_{2}=3x_{3}$ building
a complete subgraph. As discussed for the complete subgraphs, see
page \pageref{exa:We-see-that} ff., it depends on the context which
variable to take to describe the other variables and thus exploit
the functional information in the data set. The PFA can be used both
in the causal case and the non-causal case. However, when using the
PFA, a decision which variables to model by which variable, sometimes
has to be made, see Remark \ref{rem:We-can-express}. The reason is
that the functional relations, which the PFA identifies, is more general
then a causality which is a special functional relation. The PFA structures
the relations within the data set and breaks the total data set down
to subsets on which decisions can be clearly made if necessary.

Two interesting use cases of the discussion of the last paragraph
is the following. The PFA alone or combined with a causal inference
method can help to build the topology of a Bayesian Network \cite{nielsen2009bayesian},
in particular in large scales. The topology is basic to calculate
the distributions of events of an observed system to infer causes
for outcomes of random experiments. A second example is generating
the topology of gene regulatory networks that are introduced in the
introduction. 

Furthermore, the PFA can analogously be applied to the linear and
non-linear Granger-causality \cite{granger1969investigating,hiemstra1994testing,tank2018neural}.
The Granger-causality is a framework to investigate if time series
can be predicted from its own history or if there is a better prediction
with the inclusion of the history of the time series of other quantities.
From the causality information a graph is generated. The PFA can be
applied to the causal information graph when we convert the causal
information graph into a dependency graph by replacing any directed
edge with an undirected edge to investigate large scale graphs for
instance to obtain extra insights into the relations. From the PFA
resulting variables all the removed variables may be modeled. While
Granger-causality is a concept explicitly for time series data, the
PFA can be applied to both time series data and to data where the
timely relation of the data points is not given as shown in the present
work. The PFA identifies the features that are functions of other
features. A functional dependency between features also includes the
case where a feature reacts time delayed to another feature which
we may call a causal relation. A feature that reacts time delayed
to another feature can be modeled as a composition of functions. This
composite function is a function with a delay function in the function's
argument modeling the time shift.

The PFA is similar to the class of methods minimizing redundancy maximizing
relevance (mRMR) \cite{peng2005feature,ebrahimpour2013maximum,radovic2017minimum,zhao2019maximum,bugata2020some,mandal2013improved}.
The basic concept of mRMR methods is to find a selection of features
that is most relevant to an output function, meaning that the information
of the feature is well suited for a prediction of the output variable,
and at the same time that the features within the selection are minimally
redundant to each other feature in the selection, meaning that the
information to predict a feature from a different feature in the selection
is small. To describe the relevance and the redundancy of features
different scores, like the mutual information, are defined. Any selection
of features can be mapped to these scores and the selection is optimized
with regard to these scores. Consequently, the mRMR methods need to
solve (non-linear) integer optimization problems. The difference of
the PFA to the mRMR methods is that the PFA does not optimize a selection
of features with respect to analytical scores. The PFA uses algebraic
information from an independency graph resulting from a binary test
if the hypothesis that features are independent of each other has
to be rejected to identify features that are functions of others.
As discussed in Remark \ref{rem:The-dependency-graph}, the graph
can be generated from any score that measures the relation of two
features. Consequently, the algebraic information of the dependency
graph can be combined with existing mRMR methods to accelerate the
solution of the corresponding integer optimization problems which
often suffer from a challenging scaling. A possible strategy can be
to apply the PFA to dissect the dependency graph to a certain degree
and then apply an mRMR method to these disjunct subgraphs. The results
from each subgraph are independent based on the used score or test
of independency. Thus the results from each subgraph can be joined
without touching the redundancy of the total set of selected features.
The aim of the PFA is to delete all the features that are functions
of other features and thus to leave only the independent features
that are the arguments of all the functions. The arguments can be
used to model the dynamics of the data set or the features corresponding
to the arguments can be further reduced by picking only these features
of which output variables are not independent. Consequently, the procedure
of the PFA framework is an alternative method to the mRMR methods
to provide possibly different features to start modeling from. Using
another set of input features can be useful if modeling from a given
set of input features seems challenging. Furthermore, the number of
features returned by the PFA can be an initial guess for an mRMR method
for the size of its selection of features in order to delimitate the
combinatorial effort. The basic difference of the PFA and an mRMR
method is in deleting all features that are a function of features
instead of solving integer optimization problems to find a selection
of features optimizing different scores.

\section*{Discussion}

Next, we discuss the experience that we made with the chi-square test
during the experiments for the presented results. In this work a chi-square
test was used to identify the independency of two random variables.
The binning of random variables, i.e. the discretization of the co-domains
of continuous random variables, had an effect on the results and thus
the strategy for the binning was crucial. The bin size influences
how well the continuous range of a random variable is approximated
by the discretized co-domain and thus how well the discretization
represents the actual distribution of values of the random variable.
On the one hand if the binning of the random variables was too fine,
i.e. there were many bins with only a little number of data points
each bin, we fell below the recommended threshold for the minimum
number of data points in a bin consisting of the cross product of
the discretized co-domains of each two random variables for the chi-square
test (see the definition of $f_{E}$ in Section \ref{sec:The-principle-variable}
for details about the cross product bin). On the other hand, if the
binning of a random variable was too coarse, that means there is only
a small number of bins for this variable, the structure of the co-domain
of the random variable was lost where e.g. more random variables were
sorted as a constant function. 

The structure of the co-domain is important for testing the relation
of two variables and the capability to predict the output function
though. The binning strategy may have an influence on the result.
However, since the machine learning models are independent of the
binning, the sufficient prediction results on the test data sets of
the machine learning models indicate that the approximation of the
continuous co-domains via the binning recovered sufficient information.
In order to find a good bin size, we recommend the following strategy
to find a bin size $\nu$ in Algorithm \ref{alg:Discretize-the-co-domain}.
Start with a small bin size. If the PFA alerts that there is a bin
consisting of the cross product of the discretized co-domains of two
random variables in the chi-square test with less than the recommended
number of data points (default is 5), then slightly increase $\nu$
until there is no alert. By the slight increments of $\nu$ as much
of the continuous structure of the co-domain of the continuous random
variables as possible can be preserved.

The chi-square test can be so much the better be applied in cases
where the co-domain of the random variables is naturally discretized.
An example could be in bioinformatics where the expression level of
genes is classified in low, medium and high and the cell state is
either pathological or physiological.

\section*{Conclusion}

In this work the  PFA was presented. The basic idea of the PFA is
to reduce a set of features by identifying features that are functions
of other features and sort the functions out such that only these
features are left that are the arguments of the functions. If necessary
the features that are functions can be modeled from the arguments.
The basic ingredients for the PFA is a dependency graph that is built
from information about the relation of two features. In the dependency
graph functions of features generate special structures that are detected
with a minimal cut algorithm. Thus redundancies in the data set can
be reduced and the modeling can be focused on the relevant independent
input features.

The reduction of data sets improves their processing and opportunities
to understand the relations of the underlaying processes. In particular,
since the reduction of features takes place on the original data set
without transforming them into new composite features, the contribution
of the PFA will enhance the explainability of machine learning models.
Furthermore, mechanistic modeling will be simplified since the PFA
can provide candidates to construct input-output-interfaces via which
well analyzed submodels can be assembled together to a total model.
Moreover, the PFA will contribute to the capability of humans to learn
from unsupervised machine learning algorithms by extracting the features
whose values are relevant for the clustering of the machine learning
model.

The runtime of the PFA can be further developed by improving the minimal
cut algorithm provided by the networkx library for dissecting the
graph. Furthermore, the provided code can be parallelized in order
to process several of the  sub graphs in the PFA method at the same
time. In addition the provided Python code can be transformed to a
hardware near programming language.

\section*{Supplementary material}

We provide a Python implementation of the presented framework. The
three files are:
\begin{itemize}
\item execute\_relevant\_PFA.py
\item find\_relevant\_principle\_features.py
\item principle\_feature\_analysis.py
\end{itemize}
The implementation can be used for any data set with one output function
and can be easily extended to be used for more than one output function.
In the file execute\_relevant\_PFA.py, we enter the path of the considered
csv-file in which the data is stored as follows. For each data point
a column is used where in the first row we have the value of the output
function and in the remaining rows there are the values of each feature
for the corresponding data point. By executing execute\_relevant\_PFA.py
the left two functions are automatically run and the principle variables
are presented as numbers corresponding to the row number of the input
csv-file (number 1 is the input function, number 2 corresponds to
the first feature, ...). 

In the zip-file data\_server.zip there is the train and the test file
used for the results presented in Table \ref{tab:Statistics-for-different}
and Table \ref{tab:Number-of-metrics} in Section \ref{sec:An-application-of}.
In the zip-file data\_single\_cell.zip, we have the train and test
data to generate the results presented in Section \ref{sec:Application-of-the}.

The script classifiction.py was used to train the NN and the SVM from
the selected features encoded as a list of numbers corresponding to
the rows of the corresponding csv-file. The script is provided to
simplify the confirmability of the presented results. 

The script process\_data\_for\_mRMR.py can be used to transform the
csv-file in which the data points are stored for the PFA to a csv-file
for the mRMR method from http://home.penglab.com/proj/mRMR/. The version
that was used in this work is provided in the folder mRMR\_c\_source\_code.

A link where the supplementary material can be downloaded will be
available in a version of the presented work published in a journal.

\section*{Authors' contributions}

TB developed the PFA and implemented the PFA in Python. TB was mainly
involved in writing. LR set up the data center to create the data
for Section \ref{sec:An-application-of} and was involved in writing.
CL researched and preprocessed the data set for Section \ref{sec:Application-of-the}
and was involved in writing. PJ contributed to evaluating the PFA
presented in Section \ref{sec:An-application-of}  and was involved
in writing.

\section*{Conflicts of interests}

The authors declare no conflict of interest.

\section*{Acknowledgment}

The authors thank David Gengenbach (Universit\`{a} della Svizzera italiana),
Vincent Riesop (SAP SE), Matthias Rost (TU Berlin) and Hanna Kruppe
(TU Darmstadt) for fruitful discussions and their help making various
codes run. 

\section*{Appendix}

In the appendix, we give further technical details about how to discretize
the co-domain of our random variables in order to transform continuous
random variables into discrete random variables, i.e. variables with
a finite number of outcomes. Furthermore, we discuss requirements
such that we can apply our framework based on the chi-square test. 

Next, we give and subsequently discuss an algorithm to discretize
the co-domain of random variables, i.e. bin the values of the random
variables. We remark that this procedure works well for our data.
However, one can think of different methods that suits a certain scenario
to discretize the continuous random variables.

\begin{algorithm}[H]
\begin{enumerate}
\item Set $\nu\in\mathbb{N}$ number of minimum number a bin
\item For any random variable $x$
\begin{enumerate}
\item Determine the minimum value $m$ and the maximum value $M$ of all
measured values of $x$
\item If $M\leq m$: Skip $x$
\item If $M>m$: 
\begin{enumerate}
\item Sort the measured data points of $x$ in ascending order.
\item Go through the data points in ascending order. Determine the range
of a bin such that there are at least $\nu$ data points within the
current bin and that the value of the last data point of the current
bin is greater than the first one of the next bin.
\item If there are less then $\nu$ data points left: Join these data points
with the last bin that has at least $\nu$ data points.
\end{enumerate}
\end{enumerate}
\end{enumerate}
\caption{\label{alg:Discretize-the-co-domain}Discretize the co-domain of random
variables}

\end{algorithm}

Algorithm \ref{alg:Discretize-the-co-domain} works as follows. After
setting the parameter $\nu$, the following steps are performed for
any random variable whose co-domain we would like to discretize. We
first determine the minimum and the maximum value of the measure points
of $x$. If $M\leq m$, it means that the random variable is constant
according to our measurements. In this case, we skip the random variable
since we cannot gain information regarding the output function. If
$M>m$, then we go through the ordered data points and put the following
$\nu$ data points into one bin. If the value of the current data
point equals the value of the next data point that is not yet put
into a bin, we additionally put all the further data points into the
current bin until the value of the next data point is greater than
the value of the last data point that is part of the current bin.
Then we start to generate a new bin. If there are only less then $\nu$
data points left, which are not put into a bin yet, then we put these
data points into the last bin with at least $\nu$ data points. Below,
we explain the importance of the parameter $\nu$ and the consequences
of its value for the chi-square test.

Next, we discuss the distribution of the random variables $\chi_{ij}$
defined in (\ref{eq: def chi_ij}) for any $i\in\left\{ 1,...,k\right\} $
and $j\in\left\{ 1,...,l\right\} $ and their mutual independence.
These random variables each are approximately normally distributed
with expectation zero and variance one and mutually independent as
follows. The discussion will give us some insight into the requirements
for our data set such that the presented framework provides reliable
results.

In our setting of Section \ref{sec:The-principle-variable}, we perform
$n_{m}$ measurements and we count how often two random variables
take each a value assigned to a certain bin. We refer to two random
variables taking each a value assigned to a certain bin as the considered
event. Our model for the measurement is the following. The data points
are randomly drawn from a huge amount of measured data points. The
probability that the drawn data point has the value within the considered
bin is denoted with $p\in\left(0,1\right)$. We assume that the conditions
for our experimental setting are constant such that $p$ is a constant
for any repetition of a measurement series. Consequently, the number
of counts how often our considered event happens within $n_{m}$ measurements
is the expectation of a Bernoulli experiment with the binary Bernoulli
variable which models the outcome of a measurement that our considered
event happens or not. The number of counts is a random variable denoted
with $v$. The probability that we have $m_{v}$ counts of our considered
event within a period of measurement is given by the Binomial distribution
\begin{equation}
P_{B}\left(m_{v}\right)=\left(\begin{array}{c}
n_{m}\\
m_{v}
\end{array}\right)p^{m_{v}}\left(1-p\right)^{n_{m}-m_{v}}.\label{eq:bern expect}
\end{equation}
Further, we have the following relation
\begin{equation}
n_{m}p=\lambda\label{eq:np=00003Dlambda}
\end{equation}
where $\lambda$ is the expectation of counts of our considered event
within $n_{m}$ measurements. 

Since in our scenarios we usually have $m_{v}$, $\lambda$ and not
$p$, we use (\ref{eq:np=00003Dlambda}) to replace $p$ in (\ref{eq:bern expect})
by $m_{v}$ and $\lambda$. We can simplify the Bernoulli distribution
since we assume that $n_{m}$ is large compared to $m_{v}$. By applying
this assumption that $n_{m}$ is large compared to $m_{v}$, the Bernoulli
distribution can be approximated by the Poisson distribution as follows.
We have
\[
\begin{split}\lim_{n_{m}\to\infty}P_{B}\left(m_{v}\right) & =\lim_{n_{m}\to\infty}\left(\left(\begin{array}{c}
n_{m}\\
m_{v}
\end{array}\right)p^{m_{v}}\left(1-p\right)^{n_{m}-m_{v}}\right)\\
 & =\lim_{n_{m}\to\infty}\left(\frac{n_{m}!}{\left(n_{m}-m_{v}\right)!m_{v}!}\left(\frac{\lambda}{n_{m}}\right)^{m_{v}}\left(1-\frac{\lambda}{n_{m}}\right)^{n_{m}}\left(1-\frac{\lambda}{n_{m}}\right)^{-m_{v}}\right)\\
 & =\lim_{n_{m}\to\infty}\left(\frac{n_{m}!}{n_{m}^{m_{v}}\left(n_{m}-m_{v}\right)!}\left(1-\frac{\lambda}{n_{m}}\right)^{-m_{v}}\frac{\lambda^{m_{v}}}{m_{v}!}\left(1-\frac{\lambda}{n_{m}}\right)^{n_{m}}\right)\\
 & =\lim_{n_{m}\to\infty}\left(\frac{n_{m}!}{n_{m}^{m_{v}}\left(n_{m}-m_{v}\right)!}\left(1-\frac{\lambda}{n_{m}}\right)^{-m_{v}}\right)\lim_{n_{m}\to\infty}\left(\frac{\lambda^{m_{v}}}{m_{v}!}\left(1-\frac{\lambda}{n_{m}}\right)^{n_{m}}\right)\\
 & =1\cdot\frac{\lambda^{m_{v}}}{m_{v}!}\mathrm{e}^{-\lambda}
\end{split}
\]
where we use the calculation rules for the limit, see \cite[II.2 2.2 Proposition, 2.4 Proposition]{amann2005analysis}
for example, the definition of the exponential function $\lim_{n\to\infty}\left(1-\frac{\lambda}{n_{m}}\right)^{n_{m}}=\mathrm{e}^{-\lambda}$,
see \cite[III 6.23 Theorem]{amann2005analysis} for instance and the
definition of the binomial coefficient $\left(\begin{array}{c}
n_{m}\\
m_{v}
\end{array}\right)\coloneqq\frac{n_{m}!}{\left(n_{m}-m_{v}\right)!m_{v}!}$. As expected the Poisson distribution has expectation and variance
$\lambda$, see \cite[Example 1.6.4]{durrett2019probability} or \cite[1.3.4 Poisson distribution]{pinsky2010introduction}
for example. 

Now, we show that the Poisson distribution can be approximated by
a normal distribution for $\lambda$ sufficiently large and thus that
$\chi_{ij}$ for any $i\in\left\{ 1,...,k\right\} $ and $j\in\left\{ 1,...,l\right\} $
is normally distributed for $\lambda$ sufficiently large. Since the
conditions in our scenario are assumed to be constant and the data
points are randomly drawn, the following two ways to measure the counts
of our considered event are supposed to be equivalent. We can measure
the counts of our considered event drawing $n_{m}$ data points at
once. Equivalently, we can divide the $n_{m}$ measurements into several
smaller sub measurement series such that the sum of all measurements
equals $n_{m}$ as well and then sum up all the sub expectations from
any sub measurement series to obtain the expectation of the total
measurement where we perform the measurement in one part. According
to the assumption that the data points are drawn randomly, the counts
of our considered event in any sub measurement series are mutually
independent random variables following each a Poisson distribution
with corresponding expectation and variance. The corresponding mutually
independent random variables in each sub measurement series are denoted
with $v_{i}$, $i\in\left\{ 1,...,n_{v}\right\} $ where $n_{v}\in\mathbb{N}$
is the number of sub measurement series of the total measurement series.
Consequently, we can model the process of measuring the counts of
our considered event with the random variable $S=\sum_{i=1}^{n_{v}}v_{i}$
for the total measurement series.

Next, we show that $S$ is also Poisson distributed where the expectation
is the sum of all the expectations of the Poisson distributions for
the sub measurements $v_{i}$. Analogously, for the variance. We show
the Poisson distribution of $S$ by iteratively merging two Poisson
distributions. Since $v_{i}$ are mutually independent, the probability
that the sum of $v_{i}$ and $v_{j}$, $i\neq j$, equals $N_{v}\in\mathbb{N}\cup\left\{ 0\right\} $
is given by
\[
P\left(v_{i}+v_{j}=N_{v}\right)=\sum_{m_{v_{i}}=0}^{N_{v}}P_{v_{i}}\left(m_{v_{i}}\right)\cdot P_{v_{j}}\left(N_{v}-m_{v_{i}}\right)
\]
where $P_{v}$, $v\in\left\{ v_{i},v_{j}\right\} $ is the corresponding
Poisson distribution for $v_{i}$, resp., $v_{j}$ with corresponding
expectation and variance. By the virtue of the Binomial theorem, see
\cite[I 8.4 Theorem]{amann2005analysis} for example, we have, analogously
to \cite{klenke2013probability}, that the distribution of the sum
of $v_{i}$ and $v_{j}$ is also Poisson distributed with an expectation
being equal to the sum of the expectation of $P_{v_{i}}$ and $P_{v_{j}}$.
The same holds for the variance. 

Due to the reasoning of the last paragraph, we can consider a series
of measurements where our considered event occurred $\lambda$ times
as a Poisson distributed random variable decomposed of $\lambda$
random variables that are Poisson distributed with each expectation
one. For example we can choose the length of the sub measurement series
sufficiently small such that the expected occurrence of our considered
event on this sub interval equals one. 

Now, we switch to Section \ref{sec:The-principle-variable} to see
that $\chi_{ij}$ for any $i\in\left\{ 1,...,k\right\} $ and $j\in\left\{ 1,...,l\right\} $
is normally distributed for $\lambda$ sufficiently large and mutually
independent for all $i\in\left\{ 1,...,k\right\} $ and $j\in\left\{ 1,...,l\right\} $.
We have that the counts where our considered event happens, named
by $f_{O}\left(i,j\right)$ for each $i\in\left\{ 1,...,k\right\} $
and $j\in\left\{ 1,...,l\right\} $, is considered as a sum of $f_{E}\left(i,j\right)$
(which has the role of $\lambda$ in the discussion above) Poisson
distributed mutually independent random variables each with expectation
one and variance one. Consequently, we apply the Central Limit Theorem,
see \cite[Theorem 15.37]{klenke2013probability}, to obtain that $\chi_{ij}$
for each $i\in\left\{ 1,...,k\right\} $ and $j\in\left\{ 1,...,l\right\} $
is (approximately) normally distributed as follows. If $f_{E}\left(i,j\right)$
is sufficiently large, we have that $\chi_{ij}$ is normally distributed
with expectation zero and variance one for each $i\in\left\{ 1,...,k\right\} $
and $j\in\left\{ 1,...,l\right\} $. Next, we consider the mutual
independence of the $\chi_{ij}$ for all $i\in\left\{ 1,...,k\right\} $
and $j\in\left\{ 1,...,l\right\} $. Analogously to the reasoning
above for the random variable $v$ describing that a data point is
in a certain bin, we can define such a random variable for any bin
to which the values of the considered two random variables are assigned
to. Repeating the measurement series several times under constant
conditions, the variable $f_{O}\left(i,j\right)$ is the random variable
modeling the counts that the value of a data point is assigned to
the bin parametrized with $i\in\left\{ 1,...,k\right\} $ and $j\in\left\{ 1,...,l\right\} $.
Since the data points are randomly drawn from the huge amount of data
points with constant experimental settings, the expectation and variance
$f_{E}\left(i,j\right)$ is assumed constant in any sweep of a measurement
series for any $i\in\left\{ 1,...,k\right\} $ and $j\in\left\{ 1,...,l\right\} $.
As a consequence the random variables $\chi_{ij}$ are mutually independent
for all $i\in\left\{ 1,...,k\right\} $ and $j\in\left\{ 1,...,l\right\} $.

According to the reasoning of the last paragraph, we can apply the
chi-square test to $\chi^{2}$ defined in (\ref{eq:chi-square def})
as follows. Since the random variables $\chi_{ij}$ are mutually independent
and (approximately) normally distributed, the random variable $\chi^{2}$
is chi-square distributed as a sum of normally distributed independent
random variables, see \cite[Chapter 18]{cramer1999mathematical} for
example. From the chi-square distribution, the probability to obtain
a $\chi^{2}$ value greater or equal than the given value of $\chi^{2}$
can be calculated. A probability for obtaining $\chi^{2}$ values
greater or equal than the given value of $\chi^{2}$ of the performed
measurement series below a predefined threshold (level of significance)
indicates that the obtained $\chi^{2}$ values, resulting from the
deviations of $f_{O}\left(i,j\right)$ and $f_{E}\left(i,j\right)$
for one $i\in\left\{ 1,...,k\right\} $ and $j\in\left\{ 1,...,l\right\} $,
is unlikely. Although all the assumptions are fulfilled, in particular
that the considered random variables are independent, an unlikely
outcome of a random experiment can be coincidence and thus possible.
In our case however, if the probability to obtain a $\chi^{2}$ value
greater or equal than the given $\chi^{2}$ value is below a predefined
threshold, we assume that one assumption must be violated. Since the
conditions during our experiments can be assumed constant and the
data points are randomly drawn, we assume that most likely $\chi_{ij}$
is not normally distributed with zero expectation for one $i\in\left\{ 1,...,k\right\} $
and $j\in\left\{ 1,...,l\right\} $. Consequently, $f_{E}\left(i,j\right)$
is not the expectation of $f_{O}\left(i,j\right)$ for one $i\in\left\{ 1,...,k\right\} $
and $j\in\left\{ 1,...,l\right\} $ meaning we reject the hypothesis
that the corresponding random variables are independent since (\ref{eq: independent random varaiables})
is not fulfilled for one $i\in\left\{ 1,...,k\right\} $ and $j\in\left\{ 1,...,l\right\} $
within our statistical tolerances. 
\begin{rem}
\label{rem:We-remark-that}We remark that in literature it is a common
recommendation to have five data points in a bin \cite{mchugh2013chi},
i.e. $f_{E}\left(i,j\right)\}\geq5$ for each $i\in\left\{ 1,...,k\right\} $
and $j\in\left\{ 1,...,l\right\} $, to consider $f_{E}$ as sufficiently
large such that the corresponding Poisson distribution is approximately
a normal distribution. By increasing the parameter $\nu$ in Algorithm
\ref{alg:Discretize-the-co-domain}, we can increase the number of
data points in a bin, i.e. $f_{E}\left(i,j\right)\}$ is increased
for each $i\in\left\{ 1,...,k\right\} $ and $j\in\left\{ 1,...,l\right\} $,
until the condition that $f_{E}\left(i,j\right)\}\geq5$ for each
$i\in\left\{ 1,...,k\right\} $ and $j\in\left\{ 1,...,l\right\} $
is fulfilled or any other desired lower bound for $f_{E}\left(i,j\right)$.
\end{rem}


\begin{thebibliography}{10}

\bibitem{amann2005analysis}
Herbert Amann and Joachim Escher.
\newblock {\em Analysis I}.
\newblock Springer, 2005.

\bibitem{billings2013nonlinear}
Stephen~A Billings.
\newblock {\em Nonlinear system identification: NARMAX methods in the time,
  frequency, and spatio-temporal domains}.
\newblock John Wiley \& Sons, 2013.

\bibitem{bugata2020some}
Peter Bugata and Peter Drotar.
\newblock On some aspects of minimum redundancy maximum relevance feature
  selection.
\newblock {\em Science China Information Sciences}, 63(1):1--15, 2020.

\bibitem{butler2018integrating}
Andrew Butler, Paul Hoffman, Peter Smibert, Efthymia Papalexi, and Rahul
  Satija.
\newblock Integrating single-cell transcriptomic data across different
  conditions, technologies, and species.
\newblock {\em Nature biotechnology}, 36(5):411--420, 2018.

\bibitem{chen2019single}
Geng Chen, Baitang Ning, and Tieliu Shi.
\newblock Single-cell rna-seq technologies and related computational data
  analysis.
\newblock {\em Frontiers in genetics}, 10:317, 2019.

\bibitem{chen2019outage}
Yujun Chen, Xian Yang, Qingwei Lin, Hongyu Zhang, Feng Gao, Zhangwei Xu,
  Yingnong Dang, Dongmei Zhang, Hang Dong, Yong Xu, et~al.
\newblock Outage prediction and diagnosis for cloud service systems.
\newblock In {\em The World Wide Web Conference}, pages 2659--2665, 2019.

\bibitem{colombo2012learning}
Diego Colombo, Marloes~H Maathuis, Markus Kalisch, and Thomas~S Richardson.
\newblock Learning high-dimensional directed acyclic graphs with latent and
  selection variables.
\newblock {\em The Annals of Statistics}, pages 294--321, 2012.

\bibitem{cramer1999mathematical}
Harald Cram{\'e}r.
\newblock {\em Mathematical methods of statistics}, volume~43.
\newblock Princeton university press, 1999.

\bibitem{datta2016algorithmic}
Anupam Datta, Shayak Sen, and Yair Zick.
\newblock Algorithmic transparency via quantitative input influence: Theory and
  experiments with learning systems.
\newblock In {\em 2016 IEEE symposium on security and privacy (SP)}, pages
  598--617. IEEE, 2016.

\bibitem{di2007dynamic}
Alessandro Di~Cara, Abhishek Garg, Giovanni De~Micheli, Ioannis Xenarios, and
  Luis Mendoza.
\newblock Dynamic simulation of regulatory networks using squad.
\newblock {\em BMC bioinformatics}, 8(1):462, 2007.

\bibitem{durrett2019probability}
Rick Durrett.
\newblock {\em Probability: theory and examples}, volume~49.
\newblock Cambridge university press, 2019.

\bibitem{ebrahimpour2013maximum}
Morva Ebrahimpour, Hamid Mahmoodian, and Rahim Ghayour.
\newblock Maximum correlation minimum redundancy in weighted gene selection.
\newblock In {\em 2013 International Conference on Electronics, Computer and
  Computation (ICECCO)}, pages 44--47. IEEE, 2013.

\bibitem{esfahanian2013connectivity}
Abdol-Hossein Esfahanian.
\newblock Connectivity algorithms.
\newblock In {\em Topics in structural graph theory}, pages 268--281. Cambridge
  University Press, 2013.

\bibitem{granger1969investigating}
Clive~WJ Granger.
\newblock Investigating causal relations by econometric models and
  cross-spectral methods.
\newblock {\em Econometrica: journal of the Econometric Society}, pages
  424--438, 1969.

\bibitem{greenwood1996guide}
Priscilla~E Greenwood and Michael~S Nikulin.
\newblock {\em A guide to chi-squared testing}, volume 280.
\newblock John Wiley \& Sons, 1996.

\bibitem{han2011data}
Jiawei Han, Jian Pei, and Micheline Kamber.
\newblock {\em Data mining: concepts and techniques}.
\newblock Elsevier, 2011.

\bibitem{hiemstra1994testing}
Craig Hiemstra and Jonathan~D Jones.
\newblock Testing for linear and nonlinear granger causality in the stock
  price-volume relation.
\newblock {\em The Journal of Finance}, 49(5):1639--1664, 1994.

\bibitem{jolliffe2002principal}
I.T. Jolliffe.
\newblock {\em Principal Component Analysis}.
\newblock Springer Series in Statistics. Springer, 2002.

\bibitem{karl2013jimena}
Stefan Karl and Thomas Dandekar.
\newblock Jimena: efficient computing and system state identification for
  genetic regulatory networks.
\newblock {\em BMC bioinformatics}, 14(1):1--11, 2013.

\bibitem{klenke2013probability}
Achim Klenke.
\newblock {\em Probability theory: a comprehensive course}.
\newblock Springer Science \& Business Media, 2013.

\bibitem{kobak2019art}
Dmitry Kobak and Philipp Berens.
\newblock The art of using t-sne for single-cell transcriptomics.
\newblock {\em Nature communications}, 10(1):1--14, 2019.

\bibitem{levy2020predictive}
Sebastien Levy, Randolph Yao, Youjiang Wu, Yingnong Dang, Peng Huang, Zheng Mu,
  Pu~Zhao, Tarun Ramani, Naga Govindaraju, Xukun Li, et~al.
\newblock Predictive and adaptive failure mitigation to avert production cloud
  {VM} interruptions.
\newblock In {\em 14th {USENIX} Symposium on Operating Systems Design and
  Implementation ({OSDI} 20)}, pages 1155--1170, 2020.

\bibitem{lipovetsky2001analysis}
Stan Lipovetsky and Michael Conklin.
\newblock Analysis of regression in game theory approach.
\newblock {\em Applied Stochastic Models in Business and Industry},
  17(4):319--330, 2001.

\bibitem{lopez2013randomized}
David Lopez-Paz, Philipp Hennig, and Bernhard Sch{\"o}lkopf.
\newblock The randomized dependence coefficient.
\newblock In {\em Advances in neural information processing systems}, pages
  1--9, 2013.

\bibitem{lundberg2017unified}
Scott~M Lundberg and Su-In Lee.
\newblock A unified approach to interpreting model predictions.
\newblock In {\em Advances in neural information processing systems}, pages
  4765--4774, 2017.

\bibitem{mandal2013improved}
Monalisa Mandal and Anirban Mukhopadhyay.
\newblock An improved minimum redundancy maximum relevance approach for feature
  selection in gene expression data.
\newblock {\em Procedia Technology}, 10:20--27, 2013.

\bibitem{mchugh2013chi}
Mary~L McHugh.
\newblock The chi-square test of independence.
\newblock {\em Biochemia medica: Biochemia medica}, 23(2):143--149, 2013.

\bibitem{nielsen2009bayesian}
Thomas~Dyhre Nielsen and Finn~Verner Jensen.
\newblock {\em Bayesian networks and decision graphs}.
\newblock Springer Science \& Business Media, 2009.

\bibitem{papoulis2002probability}
Athanasios Papoulis and S~Unnikrishna Pillai.
\newblock {\em Probability, random variables, and stochastic processes}.
\newblock Tata McGraw-Hill Education, 2002.

\bibitem{pawar2020assessment}
Karishma Pawar and Vahida~Z Attar.
\newblock Assessment of autoencoder architectures for data representation.
\newblock In {\em Deep Learning: Concepts and Architectures}, pages 101--132.
  Springer, 2020.

\bibitem{peng2005feature}
Hanchuan Peng, Fuhui Long, and Chris Ding.
\newblock Feature selection based on mutual information criteria of
  max-dependency, max-relevance, and min-redundancy.
\newblock {\em IEEE Transactions on pattern analysis and machine intelligence},
  27(8):1226--1238, 2005.

\bibitem{picot2012flow}
Julien Picot, Coralie~L Guerin, Caroline Le~Van~Kim, and Chantal~M Boulanger.
\newblock Flow cytometry: retrospective, fundamentals and recent
  instrumentation.
\newblock {\em Cytotechnology}, 64(2):109--130, 2012.

\bibitem{pinsky2010introduction}
Mark Pinsky and Samuel Karlin.
\newblock {\em An introduction to stochastic modeling}.
\newblock Academic press, 2010.

\bibitem{radovic2017minimum}
Milos Radovic, Mohamed Ghalwash, Nenad Filipovic, and Zoran Obradovic.
\newblock Minimum redundancy maximum relevance feature selection approach for
  temporal gene expression data.
\newblock {\em BMC bioinformatics}, 18(1):1--14, 2017.

\bibitem{ribeiro2016should}
Marco~Tulio Ribeiro, Sameer Singh, and Carlos Guestrin.
\newblock " why should i trust you?" explaining the predictions of any
  classifier.
\newblock In {\em Proceedings of the 22nd ACM SIGKDD international conference
  on knowledge discovery and data mining}, pages 1135--1144, 2016.

\bibitem{samek2019explainable}
Wojciech Samek, Gr{\'e}goire Montavon, Andrea Vedaldi, Lars~Kai Hansen, and
  Klaus-Robert M{\"u}ller.
\newblock {\em Explainable AI: interpreting, explaining and visualizing deep
  learning}, volume 11700.
\newblock Springer Nature, 2019.

\bibitem{shevlyakov2016robust}
Georgy~L Shevlyakov and Hannu Oja.
\newblock {\em Robust correlation: Theory and applications}, volume~3.
\newblock John Wiley \& Sons, 2016.

\bibitem{shrikumar2017learning}
Avanti Shrikumar, Peyton Greenside, and Anshul Kundaje.
\newblock Learning important features through propagating activation
  differences.
\newblock {\em arXiv preprint arXiv:1704.02685}, 2017.

\bibitem{solomonoff2009information}
Ray~J Solomonoff, Frank Emmert-Streib, and Matthias Dehmer.
\newblock Information theory and statistical learning.
\newblock 2009.

\bibitem{vstrumbelj2014explaining}
Erik {\v{S}}trumbelj and Igor Kononenko.
\newblock Explaining prediction models and individual predictions with feature
  contributions.
\newblock {\em Knowledge and information systems}, 41(3):647--665, 2014.

\bibitem{stuart2019integrative}
Tim Stuart and Rahul Satija.
\newblock Integrative single-cell analysis.
\newblock {\em Nature Reviews Genetics}, 20(5):257--272, 2019.

\bibitem{tang2019single}
Xiaoning Tang, Yongmei Huang, Jinli Lei, Hui Luo, and Xiao Zhu.
\newblock The single-cell sequencing: new developments and medical
  applications.
\newblock {\em Cell \& Bioscience}, 9(1):53, 2019.

\bibitem{tank2018neural}
Alex Tank, Ian Covert, Nicholas Foti, Ali Shojaie, and Emily Fox.
\newblock Neural granger causality for nonlinear time series.
\newblock {\em arXiv preprint arXiv:1802.05842}, 2018.

\bibitem{wilhelm2009rna}
Brian~T Wilhelm and Josette-Ren{\'e}e Landry.
\newblock {RNA}-seq - quantitative measurement of expression through massively
  parallel rna-sequencing.
\newblock {\em Methods}, 48(3):249--257, 2009.

\bibitem{zhao2019maximum}
Zhenyu Zhao, Radhika Anand, and Mallory Wang.
\newblock Maximum relevance and minimum redundancy feature selection methods
  for a marketing machine learning platform.
\newblock In {\em 2019 IEEE International Conference on Data Science and
  Advanced Analytics (DSAA)}, pages 442--452. IEEE, 2019.

\bibitem{zhou2020visualization}
Bo~Zhou and Wenfei Jin.
\newblock Visualization of single cell rna-seq data using t-sne in r.
\newblock In {\em Stem Cell Transcriptional Networks}, pages 159--167.
  Springer, 2020.

\bibitem{zwillinger1999crc}
Daniel Zwillinger and Stephen Kokoska.
\newblock {\em CRC standard probability and statistics tables and formulae}.
\newblock Crc Press, 1999.

\end{thebibliography}
\end{document}